\documentclass[%
 reprint,
 superscriptaddress,
 amsmath,amssymb,
 aps,
 prl,
]{revtex4-2}

\usepackage{physics}
\usepackage{graphicx}
\usepackage{dcolumn}
\usepackage{bm}
\usepackage{booktabs} 
\usepackage{xcolor}
\definecolor{deepblue}{RGB}{8,48,107} 
\usepackage{hyperref}
\hypersetup{colorlinks=true,allcolors=deepblue}

\begin{document}

\author{Jonas H{\"a}nseroth}
\email{jonas.haenseroth@tu-ilmenau.de}
\affiliation{Theoretical Solid State Physics, Institute of Physics, Technische Universität Ilmenau, 98693 Ilmenau, Germany}

\author{Aaron Fl{\"o}totto}
\affiliation{Theoretical Solid State Physics, Institute of Physics, Technische Universität Ilmenau, 98693 Ilmenau, Germany}

\author{Christian Dre{\ss}ler}
\affiliation{Theoretical Solid State Physics, Institute of Physics, Technische Universität Ilmenau, 98693 Ilmenau, Germany}

\title{Universal Interatomic Potentials as Configuration-Space Generators for One-Shot and Iterative Fine-Tuning of Ab Initio-Accurate Material-Specific Models}

\date{\today}

\begin{abstract}
Universal machine-learning interatomic potentials (MLIPs) are rapidly becoming general-purpose tools for atomistic simulation, but their role in quantitative materials modeling when reactive events are involved remains unsettled.
We compare five universal MLIPs across seven chemically diverse systems and find that strong performance on standard benchmarks does not guarantee accurate predictions of target observables.
In particular, zero-shot models do not reliably reproduce reactive, transport, or high-barrier processes, exemplified here in particular by the sulfur-vacancy jump in MoS$_2$.
We therefore propose a practical alternative: universal MLIPs are used to generate long molecular dynamics trajectories, the resulting configurations are sub-sampled and relabeled with DFT, and material-specific MLIPs are subsequently trained or fine-tuned on the resulting first-principles datasets.
This workflow converts universal models into efficient configuration-space generators while retaining ab initio reference labels for training.
Across the tested systems, $2{,}000$ DFT-recalculated structures are often sufficient to obtain accurate fine-tuned or trained-from-scratch models.
For the most challenging case, iterative self-training progressively refines the sampled configuration space and recovers the DFT MoS$_2$ potential energy profile with only $600$ first-principles calculations in total.
The resulting workflow enables the generation of $1$~ns ab initio-quality trajectories  - including training data generation and model creation - within three days.
\end{abstract}

\maketitle


Atomistic simulations are central to understanding and designing functional materials, but they remain limited by the long-standing trade-off between accuracy and computational cost.
Ab initio molecular dynamics (AIMD), most commonly based on density functional theory (DFT), provides access to chemical bonding and reaction mechanisms without empirical parameterization, but is typically restricted to comparatively small systems and short timescales \cite{marx2000ab,tuckerman2002ab,iftimie2005ab}.
Classical force fields extend simulations to much larger length and timescales, but their accuracy and transferability are limited by the fixed functional form and the domain in which they were parameterized \cite{plimpton1995computational,sutmann2002classical,brooks2021classical}.
This accuracy-efficiency dilemma is particularly restrictive for processes such as ion transport, defect migration, hydrogen-bond rearrangements, solvation dynamics, and bond breaking, where long trajectories are needed but the relevant physics is controlled by subtle changes in the potential energy surface \cite{grunert2025,dressler2020effect,qaisrani2025bridging,kirsch2022atomistic}.

Machine-learning interatomic potentials (MLIPs) have emerged as a powerful route to bridge this gap by learning DFT-quality potential energy surfaces from reference calculations while retaining a computational cost closer to classical molecular dynamics \cite{kabylda2025molecular,poltavsky2025crash,pravsnikar2024machine,wang2020machine}.
Early neural-network potentials and Gaussian approximation potentials demonstrated that quantum-mechanical energies and forces can be represented accurately from data \cite{behler2007,bartok2010,friederich2021}.
More recently, graph neural networks, equivariant architectures, and symmetry-preserving representations have substantially improved the accuracy, stability, and transferability of MLIPs across chemical environments \cite{thomas2018tensor,batzner20223,mace_1,mace_2,unke2021,reiser2022,grace_1,drautz2019,kabylda2025molecular}.
As a result, MLIPs are increasingly used to obtain nanosecond-scale trajectories with ab initio accuracy for materials and molecular systems that are inaccessible to direct AIMD.

The most recent development in this field is the emergence of universal, or foundation, MLIPs.
Instead of training a potential for a single material or chemical system, these models are pretrained on large and chemically diverse databases containing millions of DFT-labeled structures \cite{jacobs2025practical,mace_mp,mattersim,orbv3,grace_2}.
Universal MLIPs therefore offer the appealing possibility of using a single pretrained model as a general-purpose simulation engine for a broad range of materials.

However, the central question is not only whether universal MLIPs perform well on benchmark metrics, but whether they can reproduce the specific observables that motivate a simulation \cite{matbench,chiang2025mlip}.
Many scientifically relevant observables depend on rare or reactive events from the high-barrier regions of configuration space rather than on equilibrium structures alone \cite{grunert2025,haenseroth2026htscreening,flototto2026large}.
Examples include diffusion barriers, proton-transfer pathways, and defect migration events \cite{haenseroth2026atk}.
Recent studies have therefore shown that, despite their broad applicability, universal models often require material-specific adaptation before quantitative accuracy can be expected in specialized applications \cite{grunert2025,flototto2026large,hanseroth2025optimizing,weiske2025statistics,chen2025high,liu2025fine,kaur2025data,Deng2025,haenseroth2026revealing}.

Fine-tuning provides one route to this material-specific accuracy \cite{tompa2026fine}.
By adapting a pretrained model to a smaller DFT-labeled dataset of the target system, one can correct the potential energy surface in the region relevant to the material of interest \cite{grunert2025,flototto2026large,hanseroth2025optimizing,weiske2025statistics,chen2025high,liu2025fine,kaur2025data,haenseroth2026revealing,radova2025fine}.

Several specialized fine-tuning strategies can further control how much of the pretrained model is adapted.
In frozen-layer fine-tuning, only selected parts of the network, often the final interaction or readout layers, are optimized while the remaining parameters are fixed \cite{radova2025fine}.
This can reduce the risk of overfitting and preserve the general chemical knowledge of the foundation model, but may limit flexibility when the target system differs strongly from the pretraining distribution.
Low-rank adaptation (LoRA) approaches instead add a small number of trainable low-rank parameters to the pretrained model, enabling efficient adaptation with reduced memory requirements and fewer trainable weights \cite{hu2022lora, grandel2026parameter}.
Multi-head replay fine-tuning keeps a shared representation of the atomic environments while assigning separate output heads to different datasets or tasks \cite{mace_1,sevennet_2,beck2025multihead}.
During replay, previously learned datasets are included again in the training process, which helps retain earlier information while the shared model is adapted to newly generated configurations.

In this work, we investigate how universal MLIPs can be used to obtain material-specific, ab initio-accurate models for particularly demanding systems in which their conventional role as weight initialization for fine-tuning may be insufficient or may require a large amount of ab initio fine-tuning data.
Rather than treating them primarily as final production potentials, we use them as efficient configuration-space generators for the construction of material-specific MLIPs.
In this workflow, a universal model is first used to generate long molecular dynamics trajectories at low cost.
The trajectory is then sub-sampled, and only the selected structures are recalculated with DFT.
The final material-specific potential is trained or fine-tuned exclusively on these DFT-labeled configurations.
This separation of sampling and labeling is essential: errors in the universal model energies and forces do not directly enter the training labels, while the universal model still provides inexpensive access to broad and physically meaningful regions of configuration space.
The approach therefore exploits the transferability of universal MLIPs without relying on their zero-shot predictions to be quantitatively correct.

We systematically assess this strategy across seven chemically distinct systems that span different bonding motifs, phases, and dynamical processes: cesium dihydrogen phosphate (CDP) and its derivative Cs$_7$(H$_4$PO$_4$)(H$_2$PO$_4$)$_8$ (CPP) \cite{struct_cdp, struct_cpp, dressler2023coexistence}, L-pyroglutamate-ammonium (L-Pyro) \cite{stephens2021short,miron2023carbonyl,qaisrani2025acid,qaisrani2025bridging}, solvated phenol (PhOH), aqueous potassium hydroxide (KOH) \cite{haenseroth2025ohlmc}, crystalline Li$_{13}$Si$_4$ \cite{zeilinger2013revision,kirsch2025li+,kirsch2022atomistic} and MoS$_2$ \cite{Li2018, Spetzler2024, flototto2026large} containing sulfur vacancies.
These systems probe non-equilibrium properties, rare and reactive events: proton transport, low-barrier hydrogen bonds, liquid-phase solvation, concentrated electrolytes, intermetallic bonding, and defect dynamics in a two-dimensional material.
We first evaluate the zero-shot performance of nine universal MLIPs, ranging from earlier foundation models to state-of-the-art models such as \textsc{MACE-OMAT-0} \cite{batatia2025cross}, \textsc{GRACE-2L-OAM} \cite{grace_2}, \textsc{PET-OAM-XL} \cite{pet_oam} and \textsc{PET-MAD-S-v1.5.0} \cite{pet_mad, pet_mad_15}.
We then compare material-specific models obtained by fine-tuning on datasets generated either from AIMD trajectories or from DFT-recalculated universal-model trajectories.

Together, these results establish practical guidelines for constructing material-specific MLIPs from limited DFT data.
Universal MLIPs are most reliable when used as accelerators of the dataset-generation process: they provide inexpensive access to long and diverse trajectories, while DFT remains responsible for the final reference labels.
The resulting material-specific models enable nanosecond-scale, first-principles-quality atomistic simulations within a practical computational workflow, while retaining validation against the physical observables of interest.
Furthermore, the workflow proposed here can be applied fully automatically with minimal user input and does not require system-specific procedures.

\subsection{State-of-the-art universal models}

\begin{figure}
    \centering
    \includegraphics[width=0.45\textwidth]{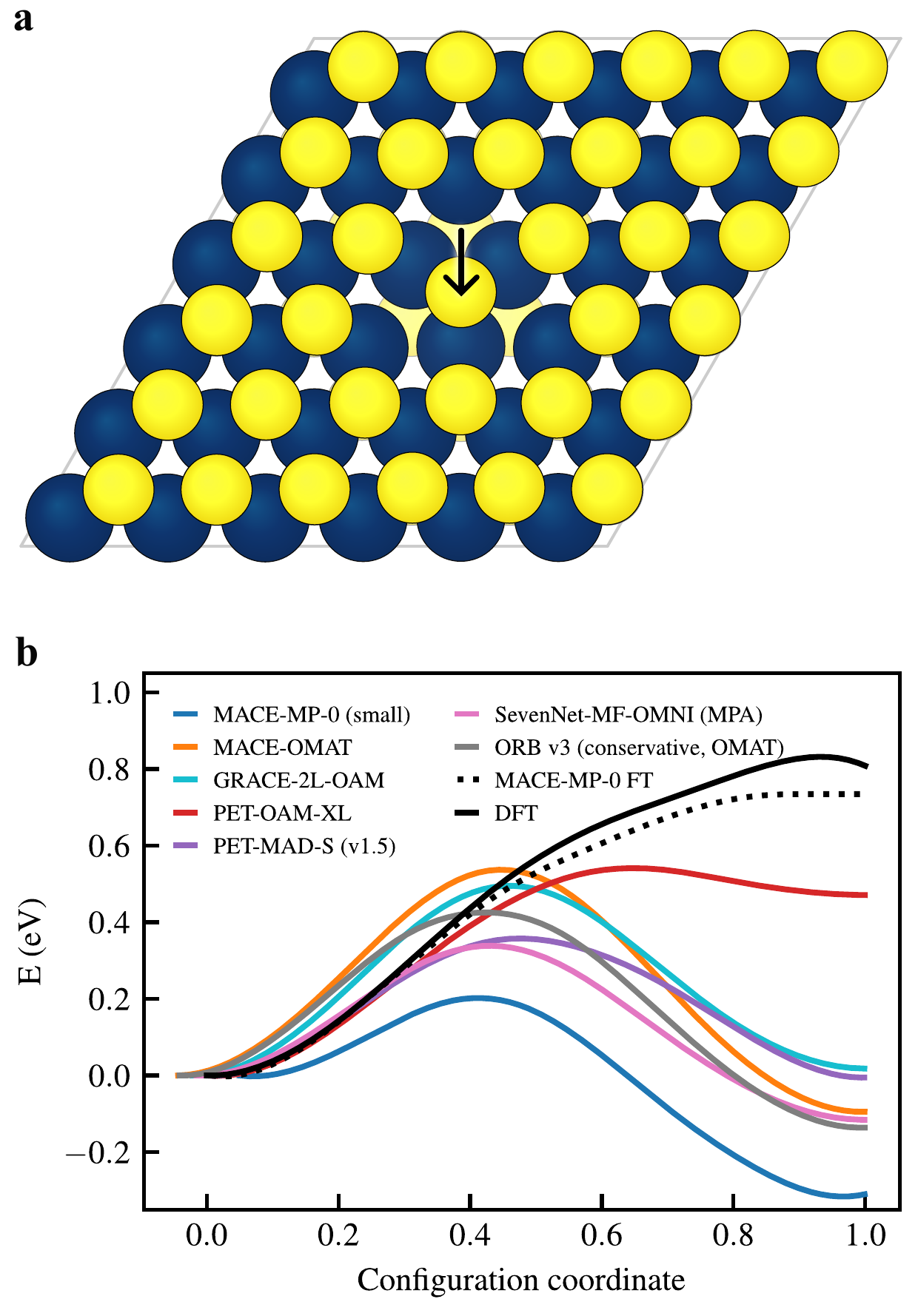}
    \caption{
    \textbf{Sulfur-vacancy jump in MoS$_2$ and the corresponding energy profile.}
    \textbf{a} Sulfur-vacancy jump in a $2$D MoS$_2$ system.
    The yellow spheres indicate sulfur atoms and the blue sphere molybdenum atoms.
    The black arrow show the sulfur-vacancy jump investigated in this work and the gray line indicates the size of the the unit.
    \textbf{b} Potential energy curves for a sulfur jump into a neighboring line of sulfur vacancies, obtained via nudge elastic band calculations with the different universal models (colored solid lines), one fine-tuned \textsc{MACE-MP-0} foundation model ($2{,}000$ equidistantly sampled AIMD frames) (black dotted line) and the DFT reference profile \cite{flototto2026large} (black solid line).
    } 
    \label{fig:modern_mos2}
\end{figure}

Universal machine-learning interatomic potentials have rapidly improved in scope, accuracy and chemical coverage \cite{pet_oam, matbench}.
The first broadly used generation of foundation MLIPs, such as \textsc{MACE-MP-0} (small) and \textsc{SevenNet-0}, was largely trained on sub-sampled \textsc{Materials Project} trajectory data \cite{mp_1,mp_2,mptrj}.
More recent universal models extend this training base to substantially larger and more diverse datasets, including structures from the \textsc{Materials Project} \cite{mp_1,mp_2,mptrj}, \textsc{Alexandria} \cite{alexandria,omat24}, \textsc{OpenMaterials2024} \cite{omat24} and \textsc{OpenMolecules2025} \cite{omol25}.
Together with larger and more advanced model architectures, this has led to a clear improvement in benchmark accuracy \cite{matbench}.
At the same time, these models have become more expensive at inference, and the central practical question remains whether their improved zero-shot performance is sufficient for target-observable-driven atomistic simulations.

We first examine this question using sulfur-vacancy dynamics in $2$D MoS$_2$ as a deliberately demanding test case (Fig.~\ref{fig:modern_mos2}), before broadening the analysis to six additional chemically diverse systems.
The sulfur-vacancy jump probes a high-energy, non-equilibrium part of the potential energy surface with a DFT barrier of about $0.8$~eV and is therefore sensitive to errors that may not be visible from equilibrium structures or small distortions alone.
This process is representative of regimes in which MLIPs are particularly useful, because long simulations are required, while also being among the most challenging cases for zero-shot performance: defect migration, ion diffusion, bond formation and breaking, and solvation in disordered phases.

The universal models predict the potential energy profile of this process with varying accuracy when compared with DFT.
Older foundation models can fail qualitatively, while newer and larger models substantially reduce the error but still do not recover the DFT reference profile with the accuracy required for reliable interpretation.
Even the best-performing zero-shot universal model, \textsc{PET-OAM-XL}, underestimates the sulfur-vacancy jump barrier by approximately $0.3$~eV.
Thus, although the newest universal MLIPs are clearly more applicable than earlier models, the improved performance does not yet guarantee ab initio-level accuracy for reactive or high-barrier material-specific observables.

In contrast, material-specific adaptation strongly improves the description of the same process.
A \textsc{MACE-MP-0} universal model fine-tuned on DFT-labeled MoS$_2$ configurations, obtained from sub-sampled AIMD trajectories, reproduces the DFT energy profile substantially more accurately than the corresponding zero-shot universal model.

To test whether this conclusion is specific to MoS$_2$ or general across chemical space, we evaluated nine universal models on seven chemically distinct systems and compared their ability to reproduce material-specific observables.
The tested models span both earlier and state-of-the-art universal MLIPs:
\textsc{MACE-MP-0} (small) \cite{mace_mp},
\textsc{GRACE-1L-OAM} \cite{grace_2},
\textsc{SevenNet-0} \cite{sevennet_1},
\textsc{MatterSim-v1.0.0-5M} \cite{mattersim},
\textsc{ORB-v2} \cite{orbv2},
\textsc{MACE-OMAT-0} (medium) \cite{batatia2025cross},
\textsc{GRACE-2L-OAM} \cite{grace_2},
\textsc{PET-OAM-XL} \cite{pet_oam},
and \textsc{PET-MAD-S-v1.5.0} \cite{pet_mad,pet_mad_15}.
The full observable-level comparison is provided in Supplementary Notes~1--9.
The corresponding force and energy errors are summarized in Supplementary Notes~10 and~11, and a detailed discussion of the accuracy of the investigated universal models is given in Supplementary Note~12.

Across the seven systems, the direct zero-shot use of universal models remains unreliable when the target quantity depends on rare or reactive events.
Some models reproduce selected observables for selected materials, but no universal model consistently reaches the desired accuracy across all tested cases.
This comparison establishes the central motivation of this work: universal MLIPs are not always reliable as final production potentials, but they may provide an efficient starting point for constructing accurate material-specific models.

\subsection{Using universal models to sample the configuration space}

\begin{figure*}
    \centering
    \includegraphics{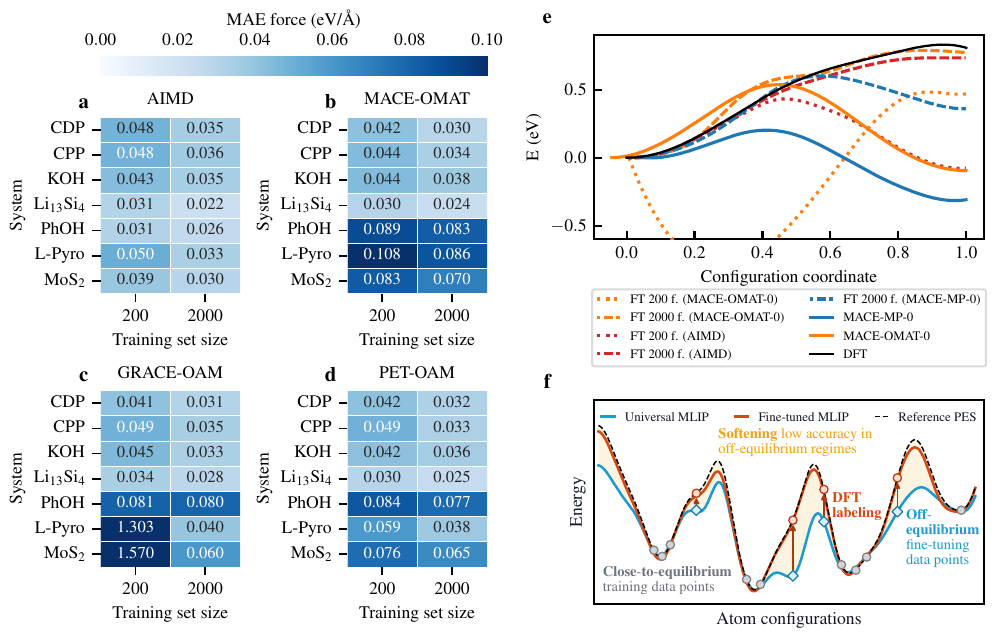}
    \caption{
   \textbf{Fine-tuning on universal model sampled structures.}
    Force mean-absolute error in eV/{\AA} of \textsc{MACE-MP-0} models fine-tuned on training sets ($200$ and $2{,}000$ data points) obtained by sub-sampled AIMD trajectories (\textbf{a}), DFT recalculated universal model structures sub-sampled from $3$~ns MD trajectories calculated with \textsc{MACE-OMAT-0} (medium) (\textbf{b}), \textsc{GRACE-2L-OAM} (\textbf{c}) and \textsc{PET-OAM-XL} (\textbf{d}) for different systems (CDP, CPP, KOH, Li$_{13}$Si$_4$, L-Pyro, MoS$_2$, and PhOH).
    \textbf{e}~potential energy curves for a sulfur jump into a neighboring line of sulfur vacancies, obtained via nudged elastic band calculations with the fine-tuned models on $200$ data points (dotted lines) and $2{,}000$ data points (dashed lines) from AIMD (red lines), \textsc{MACE-MP-0} (blue lines) and \textsc{MACE-OMAT-0} (orange lines); \textsc{MACE-MP-0} foundation model (solid blue line) and DFT (solid black line).
    \textbf{f}~Schematic illustration of the systematic softening of a two-dimensional potential energy surface by a universal MLIP.
    The DFT potential energy surface is shown as a dashed black line, while the softened potential energy surface predicted by the universal model is shown in blue.
    The orange shaded region highlights the systematic softening error of the universal model in the high-energy regime.
    Gray circles indicate configurations represented in the universal-model training data.
    Blue diamonds indicate high-energy configurations sampled during molecular dynamics with the universal model; after DFT relabeling, these configurations are shown as red circles.
    The potential energy surface obtained after fine-tuning the MLIP on the relabeled configurations is shown in red.
    } 
    \label{fig:err}
\end{figure*}

The main bottleneck in constructing material-specific MLIPs is the generation of representative first-principles reference data.
A straightforward conventional strategy, besides active learning, is to run AIMD simulations, sub-sample the resulting trajectories, and train or fine-tune a model on the selected configurations.
This approach is robust, but it becomes costly when long trajectories, large simulation cells, different concentrations or broad temperature ranges are required.
Moreover, AIMD spends first-principles effort on every simulation step, although only a small fraction of the generated structures is ultimately used for training.

We therefore use universal MLIPs as low-cost training-structure generators.
In this workflow, a state-of-the-art universal model is first used to produce long molecular dynamics trajectories of the target material.
The trajectory is then sub-sampled, and only the selected configurations are recalculated with DFT.
The final material-specific model is trained or fine-tuned exclusively on these DFT-labeled structures.
In this way, the universal model is not required to reproduce the target observable quantitatively in zero-shot mode.
Its role is instead to generate physically meaningful structures that cover relevant regions of phase space at substantially lower cost than AIMD.

This separation of sampling and labeling is essential.
Errors in the universal model energies and forces do not directly enter the training labels, because all selected configurations are recalculated with DFT.
The final accuracy is therefore controlled by the diversity and relevance of the sampled structures, together with the quality of the DFT reference labels.
Consequently, a universal model can be valuable for dataset generation even when it is not accurate enough for direct zero-shot predictions.
The level of theory used for labeling can also be changed straightforwardly, for example by recalculating the selected configurations with hybrid-functional DFT energies and forces.

Modern universal models can generate such trajectories efficiently.
For systems of approximately $1{,}000$~atoms, models such as \textsc{MACE-OMAT-0} (medium) can produce trajectories on the order of $1$~ns per day on modern high-performance GPU hardware \cite{batatia2025cross}.
With a timestep of $0.5$~fs, a $0.5$~ns trajectory contains $1{,}000{,}000$~configurations; selecting $2{,}000$~frames therefore corresponds to a stride of $500$.
The selected structures can then be recalculated with DFT as independent single-point calculations.
Because these calculations can be performed in parallel, the wall time for dataset generation can be reduced substantially compared to running a continuous AIMD trajectory.
This enables material-specific, ab initio-labeled training sets to be constructed within one to two days.

We first compare \textsc{MACE-MP-0} models fine-tuned on conventional AIMD data with models fine-tuned on DFT-recalculated structures generated by universal model trajectories (Fig.~\ref{fig:err}).
For the $2{,}000$-point datasets, trajectories generated with \textsc{MACE-OMAT-0} (medium) (Fig.~\ref{fig:err}b), \textsc{GRACE-2L-OAM} (Fig.~\ref{fig:err}c) and \textsc{PET-OAM-XL} (Fig.~\ref{fig:err}d) lead to force errors comparable to the AIMD-trained reference models (Fig.~\ref{fig:err}a) for CDP, CPP, KOH, Li$_{13}$Si$_4$ and L-Pyro.
The corresponding energy errors, shown in Supplementary Note~13, are generally higher for the universal-model trajectory datasets than for the AIMD datasets.
However, they mostly remain below $0.005$~eV/atom, compared with values mostly below $0.003$~eV/atom for the AIMD-based models.
The behavior is less favorable for MoS$_2$ and PhOH.
For these systems, models trained on structures sampled from universal-MLIP trajectories show force errors that are approximately two to three times larger than those of models trained on sub-sampled AIMD data.
The difference in energy error is also more pronounced.

The training-set size has a strong effect on the final model quality.
As expected, increasing the dataset from $200$ to $2{,}000$~DFT-labeled configurations improves the models, with the strongest influence observed for the energy accuracy.
With only $200$~data points, several models remain insufficiently accurate, and no stable fine-tuned \textsc{MACE-MP-0} model could be obtained for L-Pyro and MoS$_2$ when using DFT-labeled structures generated by \textsc{GRACE-2L-OAM}.

The MoS$_2$ sulfur-vacancy jump provides the most stringent test of whether the resulting models reproduce the target potential energy surface, because this process has a high barrier of $0.8$~eV, which renders transition-state configurations statistically rare and therefore highly sensitive to errors in the learned energy landscape.
Among the fine-tuned \textsc{MACE-MP-0} models shown in Fig.~\ref{fig:err}e, only the models fine-tuned on $2{,}000$~data points from either sub-sampled AIMD trajectories or DFT-recalculated \textsc{MACE-OMAT-0} trajectory structures reproduce the DFT energy profile with reasonable accuracy.
In contrast, a model fine-tuned on $2{,}000$~DFT-recalculated structures generated by the older and less accurate \textsc{MACE-MP-0} foundation model does not recover the DFT curve.
The $200$~data point models also fail to reproduce the correct profile.
Thus, both the quality of the structure-generating model and the size of the DFT-labeled dataset are decisive.

The corresponding sulfur-vacancy energy profiles for models fine-tuned on DFT-recalculated \textsc{GRACE-2L-OAM} and \textsc{PET-OAM-XL} trajectories are shown in Supplementary Note~14.
As for the \textsc{MACE-OMAT-0}-based datasets, the models trained on $2{,}000$~data points recover the correct qualitative and quantitative behavior.
This confirms that the target observable is considerably more sensitive than the average force error alone.
A model can reach a reasonable global error while still failing for a specific high-barrier process if the relevant configurations are not sufficiently represented in the training data.
This is particularly evident for the sulfur-vacancy jump in MoS$_2$: the force and energy errors are ensemble-averaged quantities over all atoms and configurations, whereas the NEB profile is governed by a localized displacement of a single sulfur atom along a high-energy migration pathway.
Consequently, small average errors over the full validation set do not necessarily imply an accurate description of the local potential energy surface that controls this jump process \cite{fu2022forces}.

We further evaluated whether models trained on universal-model-generated structures reproduce the material-specific observables of the remaining systems.
The results for \textsc{MACE-MP-0} models fine-tuned on \textsc{MACE-OMAT-0}-generated datasets are summarized in Supplementary Note~15, and the corresponding results for \textsc{PET-OAM-XL}-generated datasets are shown in Supplementary Note~16.
With the exception of the hydroxide ion mobility, the relevant physical properties are reproduced accurately by the fine-tuned models.
This demonstrates that universal-model-generated datasets can be sufficient for a broad range of material-specific observables, although difficult transport or reaction processes can still require additional care.
For comparison, the observables obtained from \textsc{MACE-MP-0} models fine-tuned on AIMD datasets are given in Supplementary Note~17.

We then used the DFT-labeled \textsc{MACE-OMAT-0} dataset to fine-tune additional universal models instead of \textsc{MACE-MP-0}, namely \textsc{SevenNet-0}, \textsc{MatterSim-v1.0.0-5M}, \textsc{GRACE-1L-OAM} and \textsc{ORB-v2}.
The results are shown in Supplementary Notes~18-22.
As discussed in our previous work, fine-tuning these models with a reasonably large dataset results in material-specific models with comparable performance largely independent of architecture \cite{haenseroth2026atk}.
For MoS$_2$, L-Pyro and PhOH, however, the fine-tuned models show elevated errors that are approximately two to three times larger than those of the AIMD-based fine-tuned models, similar to the fine-tuned \textsc{MACE-MP-0} models for these materials in Fig.~\ref{fig:err}b.

The ability of these different fine-tuned universal models to predict the sulfur-vacancy jump energy curve in MoS$_2$ is shown in Supplementary Note~20.
All models based on the DFT-labeled $2{,}000$-point \textsc{MACE-OMAT-0} dataset reproduce the energy profile, except for the \textsc{ORB-v2} model.
The performance of the \textsc{SevenNet-0} and \textsc{MatterSim-v1.0.0-5M} models for the material properties of the other systems is shown in Supplementary Notes~23 and~24.
These material-specific models reproduce the observables for most systems, but they do not predict the hydroxide ion diffusivity with the desired first-principles accuracy.

Finally, we considered training models from scratch instead of fine-tuning universal models.
For this comparison, \textsc{MACE} models of similar size to the \textsc{MACE-MP-0} foundation model were trained from scratch on the same datasets.
The force and energy errors are shown in Supplementary Notes~23 and~24.
For the $2{,}000$-point datasets, accurate trained-from-scratch models are obtained for CDP, CPP, KOH and Li$_{13}$Si$_4$ when using structures generated by \textsc{PET-OAM-XL}, \textsc{MACE-OMAT-0} (medium) or \textsc{GRACE-2L-OAM}.
For L-Pyro, similarly accurate models are obtained with \textsc{GRACE-2L-OAM} and \textsc{MACE-OMAT-0}.
The resulting errors are generally comparable to those of models trained from scratch on sub-sampled AIMD data and are often smaller than those of the corresponding fine-tuned \textsc{MACE-MP-0} models shown in Fig.~\ref{fig:err}a-d.
For the MoS$_2$ sulfur-vacancy jump, the energy profiles obtained with trained-from-scratch models show particularly robust behavior, as shown in Supplementary Note~25.

While using universal MLIPs to sample configurational space, we are exploiting the recently discussed systematic softening of the potential energy surface \cite{Deng2025}. 
In this context, softening refers to the tendency of universal models to underestimate energy barriers in high-energy regions of the potential energy surface, as illustrated in Fig.~\ref{fig:err}f.
This behavior likely originates from the composition of the training data, which is typically dominated by close-to-equilibrium structures. 
Although this limits the direct use of universal MLIPs for quantitative barrier predictions, it can be advantageous for dataset generation: softened barriers allow the model to visit off-equilibrium configurations more frequently during molecular dynamics. 
After these configurations are relabeled with first-principles calculations, they provide targeted reference data for training a material-specific MLIP with improved accuracy in the high-energy regions that are underrepresented in equilibrium AIMD data.

However, this strategy is not guaranteed to succeed in a single step. 
If the generated trajectory does not sample the relevant transition region sufficiently, or if the relabeled dataset still lacks configurations controlling the target observable, the resulting model may retain the same qualitative failure. 
These model failures therefore identify cases where a single-pass dataset-generation strategy is insufficient and naturally motivate an iterative refinement procedure.

\subsection{Iterative training}

\begin{figure*}
    \centering
    \includegraphics{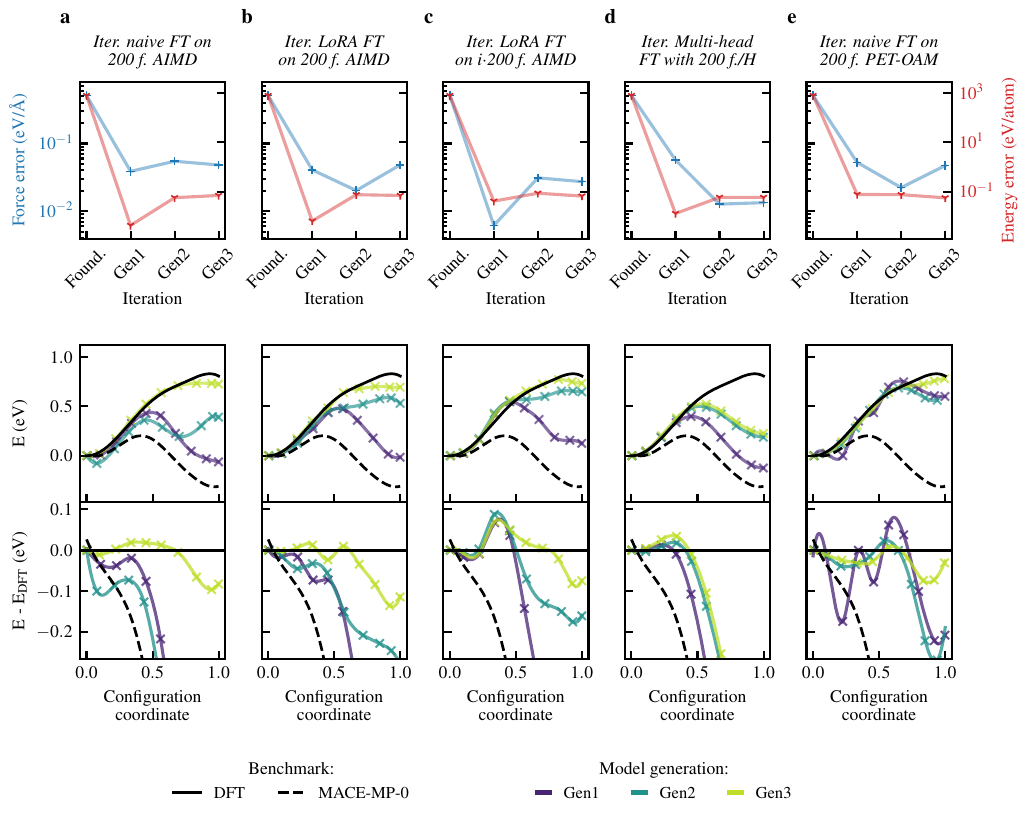}
    \caption{
    \textbf{Iterative self-training.}
    Evaluation of three generations of iteratively fine-tuned \textsc{MACE-MP-0} models for MoS$_2$.
    The upper row shows force and energy errors, the middle row shows the predicted energy profile of the sulfur-vacancy jump in MoS$_2$, and the lower row shows the deviation from the DFT reference profile.
    Different iterative fine-tuning strategies are compared:
    \textbf{a} naive fine-tuning initialized with $200$~AIMD data points and a constant dataset size of $200$ points per model;
    \textbf{b} LoRA fine-tuning initialized with $200$~AIMD data points and a constant dataset size of $200$ points per model;
    \textbf{c} LoRA fine-tuning initialized with $200$~AIMD data points and a dataset size increasing linearly by $200$ points per generation;
    \textbf{d} multi-head fine-tuning initialized with $200$~AIMD data points, using $200$ data points per generation and one head per generation;
    \textbf{e} naive fine-tuning initialized with $200$~DFT-labeled structures obtained by sub-sampling a \textsc{PET-OAM-XL} universal-model trajectory, using a constant dataset size of $200$ points per model.
    }
    \label{fig:iter}
\end{figure*}

The previous results show that DFT-recalculated universal-model trajectories provide an efficient route to accurate material-specific MLIPs, but also that a single dataset-generation step is not always sufficient.
This limitation becomes particularly important when the target observable depends on a narrow, rare, or high-energy region of configuration space.
For such cases, we investigate an iterative training workflow in which the current material-specific model is used to generate the structures for the next training iteration.

The procedure is conceptually simple.
In the first generation, a model is trained from scratch or fine-tuned using an initial DFT-labeled dataset obtained either from a sub-sampled AIMD trajectory or from a DFT-recalculated sub-sampled universal-MLIP trajectory.
This model is then used to run a molecular dynamics simulation.
The resulting trajectory is equidistantly sub-sampled, and the selected configurations are recalculated with DFT.
These newly labeled configurations are then used to fine-tune the current model and obtain the next model generation, either as a replacement for the previous dataset or in combination with data from earlier generations.
The loop is repeated until the target observable is reproduced with the desired accuracy.
In analogy to modern large-language-model workflows, this strategy can be viewed as iterative self-training or synthetic data generation \cite{syn_data_gen}; however, in the present case every newly generated structure is relabeled with first-principles calculations before it enters the training set.
This approach is particularly promising because MLIPs trained primarily on equilibrium structures show the tendency to underestimate the potential energies of transition states~\cite{Deng2025}, making such states more likely to appear in training sets sampled from MLIP-based MD simulations than from AIMD simulations.

We demonstrate this strategy for the most challenging observable considered in this work, the sulfur-vacancy jump in MoS$_2$.
The initial model is obtained by fine-tuning \textsc{MACE-MP-0} on only $200$~configurations sub-sampled from an AIMD trajectory (Fig.~\ref{fig:iter}a).
This first-generation model is then used to perform an MLIP molecular dynamics simulation of $600{,}000$~steps.
From this trajectory, $200$~configurations are equidistantly sampled and recalculated with DFT to form the training set for the next generation.
Repeating this procedure yields a third-generation model that correctly reproduces the DFT sulfur-vacancy energy profile, while requiring only $600$~first-principles single-point calculations in total and two additional MLIP molecular dynamics simulations.

This result is notable because models trained on datasets containing $600$~AIMD configurations or $600$~\textsc{MACE-MP-0}-generated configurations fail to reproduce the DFT profile, as shown in Supplementary Note~26.
Even fine-tuning on $2{,}000$~DFT-labeled configurations sampled with the older \textsc{MACE-MP-0} foundation model does not recover the correct curve (Fig.~\ref{fig:iter}e).

We further investigated whether alternative fine-tuning strategies improve the iterative workflow.
When the same iterative procedure is performed with LoRA fine-tuning instead of naive fine-tuning (Fig.~\ref{fig:iter}b), the third-generation model reaches a performance comparable to the naively fine-tuned model.
However, the second-generation LoRA model already predicts the sulfur-vacancy energy profile substantially more accurately than the corresponding naively fine-tuned model.
This indicates that LoRA fine-tuning can improve data efficiency in the early stages of iterative training, where the available DFT-labeled datasets are still small.

We then tested whether retaining data from previous generations further improves the LoRA-based workflow.
In this variant, the dataset size increases linearly by $200$~structures per generation.
For example, the second-generation model is fine-tuned not only on the $200$~newly generated configurations, but on a combined dataset containing the initial AIMD-based structures and the newly DFT-labeled structures generated from the first-generation model, resulting in $400$~data points in total.
This strategy leads to a substantially improved second-generation prediction of the sulfur-vacancy energy profile (Fig.~\ref{fig:iter}c).
The third-generation model, however, performs comparably to the iterative models trained with a constant training dataset size of $200$~points per generation.

In contrast, multi-head fine-tuning does not provide a clear additional benefit for this system.
In this strategy, each newly generated dataset is assigned to a separate head, with one head per model generation (Fig.~\ref{fig:iter}d).
Although the second-generation model improves relative to the initial model, no substantial additional improvement is observed in the third generation.
This suggests that, for the present MoS$_2$ case, separating the generations into different heads is less effective than directly adapting the shared model parameters to the newly sampled configuration space.

Finally, we tested whether the iterative workflow requires an AIMD-based initial dataset.
Starting from $200$~DFT-labeled structures obtained by sub-sampling a \textsc{PET-OAM-XL} universal-model trajectory already yields a well-performing first-generation model (Fig.~\ref{fig:iter}e).
Additional iterative fine-tuning further improves the model initially, demonstrating that the workflow can also be initialized from universal-model-generated structures.
However, during further iterations, the model accuracy decreases, as shown in Supplementary Note~27.
This behavior can be rationalized from the dataset composition of the respective training sets, shown in Supplementary Note~28, and indicates that iterative self-training must still be monitored with respect to the diversity and relevance of the generated configurations.

Together, these results show that iterative training provides a practical fallback when one-shot fine-tuning or training on AIMD- or universal-model-generated datasets does not yet reach the required target accuracy.
The strategy is particularly useful when the relevant configurations are not sufficiently represented in the initial dataset.

\section*{Conclusion and training guidelines}

This work establishes a practical workflow for constructing material-specific MLIPs with ab initio accuracy from limited first-principles data.
We systematically compared zero-shot universal models, naive fine-tuning, training-from-scratch and iterative fine-tuning across seven chemically distinct systems and several families of modern MLIPs.
The study deliberately focuses on broadly accessible strategies based on naive fine-tuning and on dataset construction through equidistant sampling of molecular dynamics trajectories.
This allows the resulting guidelines to be applied across different MLIP frameworks without relying on specialized active-learning infrastructure.

The first conclusion is that state-of-the-art universal MLIPs should generally not be treated as final production potentials for target-observable-driven simulations.
Although the newest models are substantially more accurate than earlier generations, they still fail for several observables connected to reactive events, ion mobility, defect motion or high-energy regions of the potential energy surface.
The sulfur-vacancy jump in MoS$_2$ is the clearest example: even the best zero-shot universal model underestimates the barrier by approximately $0.3$~eV.
For mechanistic interpretation, this level of error is too large because relevant observables depend exponentially on the energy barrier according to Boltzmann statistics, and ab initio-level accuracy cannot be assumed for universal models in such cases.

The second conclusion is that universal models are nevertheless highly valuable.
Their most robust role is not necessarily to replace DFT directly, but to generate diverse and physically meaningful configurations at low computational cost.
After DFT recalculation of selected frames, these structures form accurate material-specific training datasets.
Using state-of-the-art universal MLIPs as structure generators enables the exploration of nanosecond-scale trajectories before deciding which configurations should receive expensive first-principles labels.
This shifts the data-generation bottleneck from continuous AIMD to a set of independent DFT single-point calculations, which can be parallelized efficiently.

For the systems investigated here, $2{,}000$ DFT-recalculated configurations selected from trajectories generated by modern universal MLIPs are often sufficient to obtain accurate material-specific models.
When state-of-the-art universal models are used to generate the structures, approximately two thirds of the investigated cases lead to fine-tuned \textsc{MACE-MP-0} models with the desired accuracy.
Using the same datasets for training from scratch yields the desired accuracy in $61$~\% of the investigated cases.
Here, training from scratch performs on par with or better than naive fine-tuning when the dataset contains $2{,}000$ configurations.
This identifies an important advantage of universal models: even when their zero-shot predictions are not sufficiently accurate, they can strongly accelerate the generation of useful material-specific training data.

The quality of the structure-generating model is critical.
Datasets generated with older or less accurate universal MLIPs can lead to substantially worse material-specific models, even when all selected structures are relabeled with DFT.
This is because the DFT labels correct the energies and forces of the sampled configurations, but they cannot compensate for missing regions in the sampling of the configuration space.
If the generating model does not visit the relevant structures, the final material-specific model cannot learn them.
This effect is especially important for rare events and transition pathways, where average force and energy errors alone may not reveal the failure.
As universal models continue to improve, the fraction of systems for which this workflow yields accurate material-specific models is therefore expected to increase.

The results suggest the following practical workflow.
\textbf{1.} When no AIMD trajectory is available, start by generating a trajectory with a state-of-the-art universal MLIP, such as \textsc{PET-OAM-XL}, \textsc{MACE-OMAT-0} (medium) or \textsc{GRACE-2L-OAM}.
Sub-sample approximately $2{,}000$ configurations from this trajectory and recalculate these structures with parallelized DFT single-point calculations.
Use the resulting dataset to train a material-specific model from scratch or to fine-tune a suitable foundation model.
Based on the present results, \textsc{MACE} and \textsc{SevenNet} are particularly useful choices for training from scratch, while \textsc{MACE-MP-0}, \textsc{SevenNet-0} and \textsc{MatterSim-v1.0.0-5M} provide strong fine-tuning baselines.

\textbf{2.} Validate the resulting model on the physical observable of interest, not only on global force and energy errors.
For many systems, a one-shot dataset generated from a modern universal-model trajectory is sufficient.
However, observables such as the MoS$_2$ sulfur-vacancy jump show that good average errors do not always imply correct target behavior.
Target-observable validation is therefore essential, especially for high-barrier processes, transport properties and reaction mechanisms.

\textbf{3.} If the one-shot model fails, apply iterative self-training.
Use the current material-specific model to generate additional trajectories, sub-sample the configurations it visits, recalculate them with DFT and continue training or fine-tuning.
This strategy can be more efficient than simply increasing the size of the initial dataset because it exploits a potential softening of the MLIP~\cite{Deng2025}, which can automatically lead to an overrepresentation of particularly relevant transition states in the training set.
For MoS$_2$, this approach recovers the correct sulfur-vacancy energy profile using only $600$~DFT single-point calculations in total, whereas several non-iterative strategies with comparable or larger datasets fail.

Overall, universal MLIPs provide the greatest benefit when used as accelerators of the dataset-generation process.
They allow long and diverse trajectories to be generated cheaply, while DFT remains responsible for the final reference labels.
Material-specific MLIPs trained on these datasets can then reach the accuracy required for ab initio-quality simulations at a fraction of the cost of direct AIMD.
The resulting workflow enables nanosecond-scale, first-principles-quality atomistic trajectories within only a few days, while retaining a clear validation path through target observables and iterative refinement when necessary.

\section*{Methods}

\subsection*{AIMD simulations}

Reference DFT AIMD simulations were performed on CPUs using CP2K (version 2025.1) \cite{cp2k_1, cp2k_2, cp2k_3, cp2k_4, cp2k_5, cp2k_quickstep,cp2k_orb_trans} with the BLYP (KOH, L-Pyro, PhOH) \cite{blyp1,blyp2} and PBE (CDP, CPP, MoS$_2$, Li$_{13}$Si$_4$) \cite{pbe} exchange-correlation functional, GTH pseudopotentials \cite{cp2k_gth-pseudopot1, cp2k_gth-pseudopot2, cp2k_gth-pseudopot3}, DZVP-MOLOPT basis set \cite{cp2k_basis-set}, and Nos\'{e}-Hoover chain thermostats \cite{nose1,nose2,nose3} applying a timestep of $0.5$~fs.
Training data consisted of $10$, $200$, and $2{,}000$ configurations per system equidistantly sampled from AIMD trajectories at relevant temperatures: $300$~K (PhOH, L-Pyro), $333$~K (KOH), $500$~K (Li$_{13}$Si$_4$), [$510$, $540$, $585$, $620$, $660$~K] (CDP, CPP), and $1000$~K (MoS$_2$) with a stride of $10$ and $100$.

\subsection{Training and fine-tuning MLIPs}

All fine-tuning and training calculations were performed using the respective MLIP framework implementations: \textsc{MACE-torch} (version 0.3.14) \cite{mace_1,mace_2}, \textsc{GRACE tensorpotential} (version 0.5.1) \cite{grace_1,grace_2}, \textsc{SevenNet} (version 0.11.2) \cite{sevennet_1,sevennet_2}, \textsc{MatterSim} (version 1.1.2) \cite{mattersim}, and \textsc{ORB} (version 0.3.2) \cite{orbv2,orbv3} through their official releases.
For fine-tuning we employed the smaller and more efficient models of the respective MLIP frameworks: \textsc{MACE-MP-0} (small) \cite{mace_mp}, \textsc{GRACE-1L-OAM} \cite{grace_2}, \textsc{SevenNet-0} \cite{sevennet_1}, \textsc{MatterSim-v1.0.0-5M} (MatterSim-Large) \cite{mattersim} and \textsc{ORB-v2} \cite{orbv2}.
For training we used the same model sizes as the mentioned smaller foundation models, to be able to compare model performance between fine-tuning and training.
Fine-tuning and training were performed using \textsc{NVIDIA} A100 GPUs.

For production and the training set generated from DFT-recalculated foundation model structures we used more sophisticated and expensive models:
\textsc{MACE-OMAT-0} (medium) \cite{mace_mp}, \textsc{GRACE-2L-OAM} \cite{grace_2}, \textsc{SevenNet-Omni} (mpa modal) \cite{sevennet_omni}, \textsc{PET-OAM-XL} \cite{pet_oam,metatrain} and \textsc{PET-MAD-S-v1.5.0} \cite{pet_mad,pet_mad_15,metatrain}.

All errors reported in this work were computed on reference data excluded from the training and validation sets.
This test dataset was obtained by recalculating, with first-principles methods, MLIP trajectories generated using the fine-tuned foundational \textsc{MACE-MP-0} models ($2{,}000$ data points, sub-sampled AIMD trajectories, stride $100$).  

\subsection{MLIP MD and NEB calculations}

Molecular dynamics simulations were performed with \textsc{ASE} (version 3.25.0) \cite{ase} using Nos\'{e}-Hoover chain thermostats \cite{nose1, nose2, nose3} set up with \textsc{aMACEing\_toolkit} (version 0.7.0) \cite{haenseroth2026atk} for $3$~ns at $300$~K (PhOH, L-Pyro), $333$~K (KOH), $500$~K (Li$_{13}$Si$_4$), [$510$, $540$, $585$, $620$, $660$~K] (CDP, CPP), and $1000$~K (MoS$_2$) with a timestep of $0.5$~fs.
Nudged elastic band calculations were performed with \textsc{ASE} (version 3.25.0) \cite{ase}.   
MD and NEB calculations were performed using \textsc{NVIDIA} A100 GPUs.

\section*{Data Availability}

The dataset containing training sets and the models are available at \url{doi.org/10.5281/zenodo.20792778}.

\section*{Code Availability}

The used third-party codes \textsc{CP2K}, \textsc{ASE}, \textsc{MACE}, \textsc{MatterSim}, \textsc{GRACE}, \textsc{SevenNet}, \textsc{UPET} and \textsc{ORB} are available at \url{cp2k.org}, \url{ase-lib.org}, \url{github.com/acesuit/mace}, \url{github.com/microsoft/mattersim}, \url{github.com/ICAMS/grace-tensorpotential}, \url{github.com/MDIL-SNU/SevenNet}, \url{github.com/lab-cosmo/upet} and \url{github.com/orbital-materials/orb-models} respectively. 
The input-script creator used in this study is available at \url{github.com/jhaens/amaceing_toolkit}.

\section*{Acknowledgments}

We thank the staff of the Compute Center of the Technische Universität Ilmenau, especially Mr.~Henning~Schwanbeck for providing an excellent research environment. 
This work is supported by the doctoral scholarship of the German Academic Scholarship Foundation, Carl-Zeiss-Stiftung (SustEnMat, funding code: P2023-02-008), the Thüringer Aufbaubank (TAB) (KapMemLyse, grant no.~2024 FGR 0081 / 0082) and the European Social Fund Plus (ESF+).

\section*{Competing interests}

The authors declare no competing interests.

\section*{Author contributions}

J.H. and C.D.~conceived the idea, J.H.~wrote the high-throughput workflow, J.H. and A.F.~performed all calculations; J.H., A.F. and C.D.~analyzed the data; J.H. and A.F. visualized all results and J.H.~wrote the first draft of the manuscript. 
C.D.~supervised the work; all authors revised and approved the manuscript.

\bibliography{bibliography.bib}

\clearpage
\begin{figure*}
    \centering
    \includegraphics[width=8.25cm,height=4.45cm,keepaspectratio]{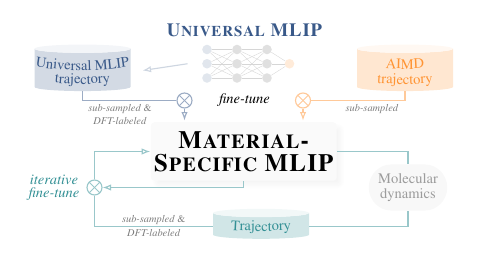}
    \caption{\textbf{TOC Graphic}}
\end{figure*}

\end{document}


\author{Jonas H{\"a}nseroth}
\email{jonas.haenseroth@tu-ilmenau.de}
\affiliation{Theoretical Solid State Physics, Institute of Physics, Technische Universität Ilmenau, 98693 Ilmenau, Germany}

\author{Aaron Fl{\"o}totto}
\affiliation{Theoretical Solid State Physics, Institute of Physics, Technische Universität Ilmenau, 98693 Ilmenau, Germany}

\author{Christian Dre{\ss}ler}
\affiliation{Theoretical Solid State Physics, Institute of Physics, Technische Universität Ilmenau, 98693 Ilmenau, Germany}

\title{Supplementary Information for "Universal Interatomic Potentials as Configuration-Space Generators for One-Shot and Iterative Fine-Tuning of Ab Initio-Accurate Material-Specific Models"}
\date{\today}

\maketitle

For more information on abbreviations, please refer to the main text, where all abbreviations are defined in detail.
Abbreviations not introduced in the main text are defined here.

\tableofcontents
\newpage
\clearpage

\section*{Supplementary Note 1: MACE-MP-0 (small) universal model - prediction of physical properties}

\begin{figure}[ht]
    \centering
    \includegraphics{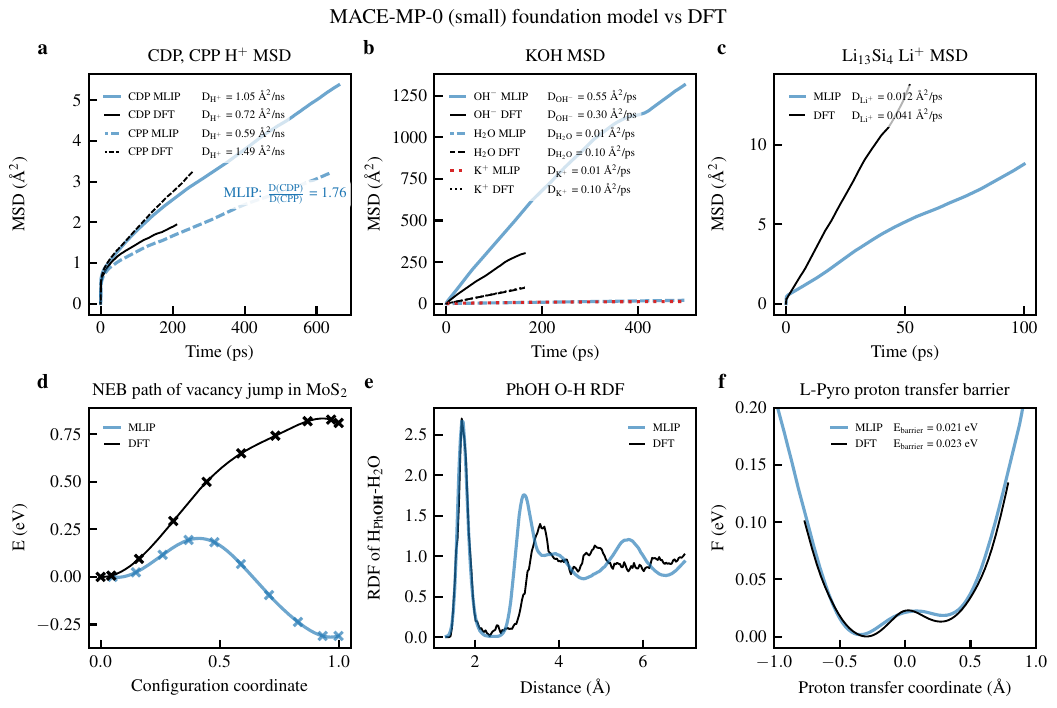}
    \caption{\textbf{MACE-MP-0 (small) universal model}.
    Comparison of different physical properties obtained with first-principles methods and the universal model: \textbf{a} CsH$_2$PO$_4$ (CDP) and Cs$_7$(H$_4$PO$_4$)(H$_2$PO$_4$)$_8$ (CPP) - proton diffusion coefficients ratios of D(CDP)/D(CPP), \textbf{b} KOH - water, water molecule and hydroxide ion mean-squared displacements and diffusion coefficients, \textbf{c} Li$_{13}$Si$_4$ - lithium ion mean-squared displacements and diffusion coefficients, \textbf{d} MoS$_2$ - potential energy curves for a sulfur jump into a neighboring line of sulfur vacancies, \textbf{e} phenol in water - (H$_2$O)$\cdots$O\textsubscript{Hydroxyl-Group} radial distribution function and \textbf{f} L-pyroglutamate-ammonium (L-Pyro) - free energy profiles along the proton transfer coordinate.
    } 
\end{figure}

\newpage
\section*{Supplementary Note 2: MatterSim-v1.0.0-5M universal model - prediction of physical properties}

\begin{figure}[ht]
    \centering
    \includegraphics{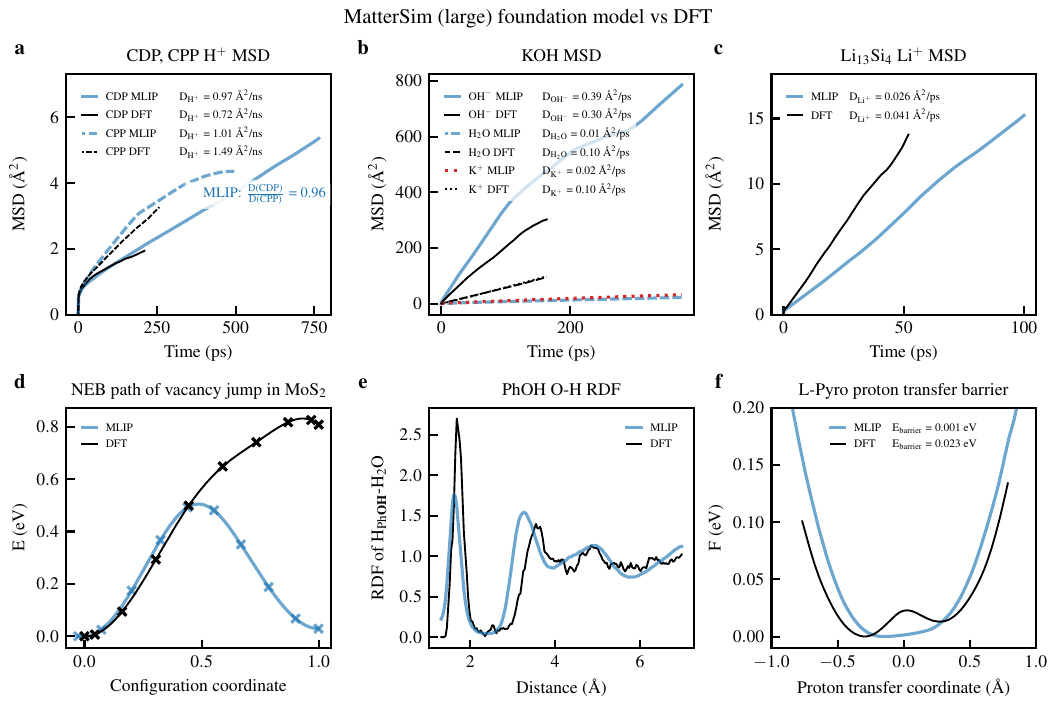}
    \caption{\textbf{MatterSim-v1.0.0-5M universal model}.
    Comparison of different physical properties obtained with first-principles methods and the universal model: \textbf{a} CsH$_2$PO$_4$ (CDP) and Cs$_7$(H$_4$PO$_4$)(H$_2$PO$_4$)$_8$ (CPP) - proton diffusion coefficients ratios of D(CDP)/D(CPP), \textbf{b} KOH - water, water molecule and hydroxide ion mean-squared displacements and diffusion coefficients, \textbf{c} Li$_{13}$Si$_4$ - lithium ion mean-squared displacements and diffusion coefficients, \textbf{d} MoS$_2$ - potential energy curves for a sulfur jump into a neighboring line of sulfur vacancies, \textbf{e} phenol in water - (H$_2$O)$\cdots$O\textsubscript{Hydroxyl-Group} radial distribution function and \textbf{f} L-pyroglutamate-ammonium (L-Pyro) - free energy profiles along the proton transfer coordinate.
    } 
\end{figure}

\newpage
\section*{Supplementary Note 3: SevenNet-0 universal model - prediction of physical properties}

\begin{figure}[ht]
    \centering
    \includegraphics{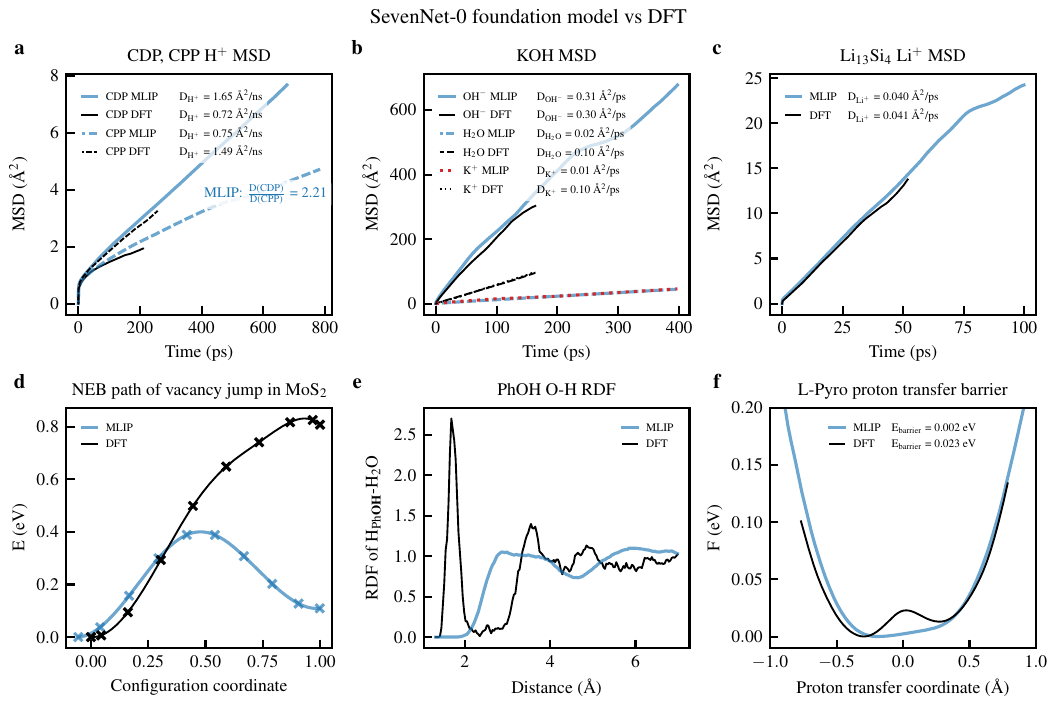}
    \caption{\textbf{SevenNet-0 (small) universal model}.
    Comparison of different physical properties obtained with first-principles methods and the universal model: \textbf{a} CsH$_2$PO$_4$ (CDP) and Cs$_7$(H$_4$PO$_4$)(H$_2$PO$_4$)$_8$ (CPP) - proton diffusion coefficients ratios of D(CDP)/D(CPP), \textbf{b} KOH - water, water molecule and hydroxide ion mean-squared displacements and diffusion coefficients, \textbf{c} Li$_{13}$Si$_4$ - lithium ion mean-squared displacements and diffusion coefficients, \textbf{d} MoS$_2$ - potential energy curves for a sulfur jump into a neighboring line of sulfur vacancies, \textbf{e} phenol in water - (H$_2$O)$\cdots$O\textsubscript{Hydroxyl-Group} radial distribution function and \textbf{f} L-pyroglutamate-ammonium (L-Pyro) - free energy profiles along the proton transfer coordinate.
    } 
\end{figure}

\newpage
\section*{Supplementary Note 4: GRACE-1L-OAM universal model - prediction of physical properties}

\begin{figure}[ht]
    \centering
    \includegraphics{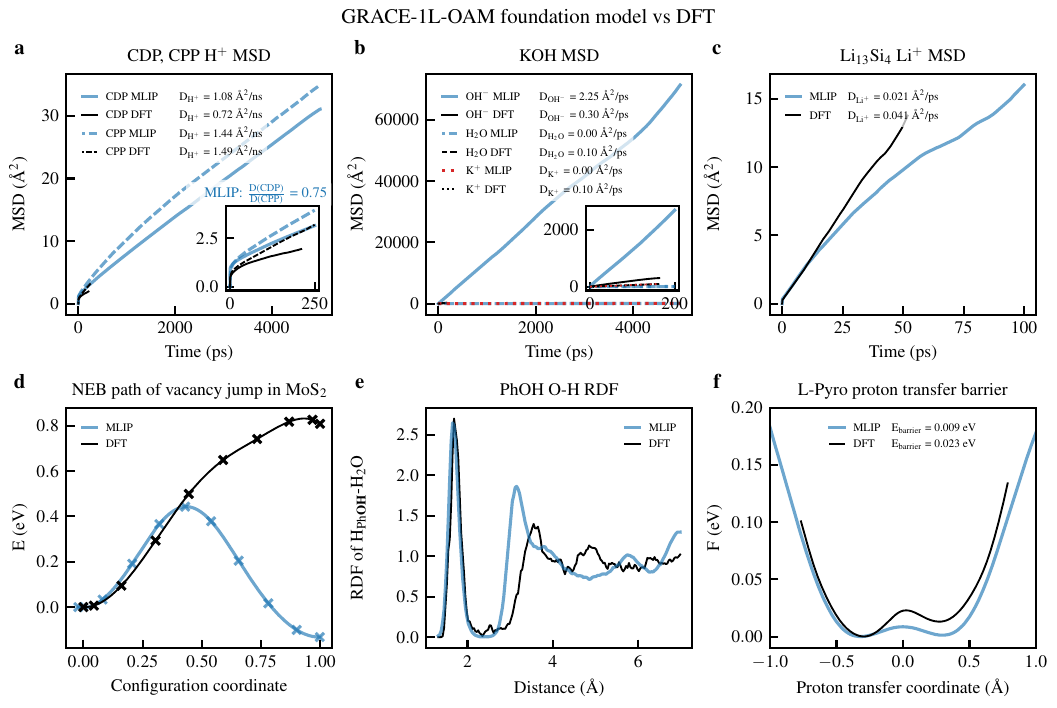}
    \caption{\textbf{GRACE-1L-OAM universal model}.
    Comparison of different physical properties obtained with first-principles methods and the universal model: \textbf{a} CsH$_2$PO$_4$ (CDP) and Cs$_7$(H$_4$PO$_4$)(H$_2$PO$_4$)$_8$ (CPP) - proton diffusion coefficients ratios of D(CDP)/D(CPP), \textbf{b} KOH - water, water molecule and hydroxide ion mean-squared displacements and diffusion coefficients, \textbf{c} Li$_{13}$Si$_4$ - lithium ion mean-squared displacements and diffusion coefficients, \textbf{d} MoS$_2$ - potential energy curves for a sulfur jump into a neighboring line of sulfur vacancies, \textbf{e} phenol in water - (H$_2$O)$\cdots$O\textsubscript{Hydroxyl-Group} radial distribution function and \textbf{f} L-pyroglutamate-ammonium (L-Pyro) - free energy profiles along the proton transfer coordinate.
    } 
\end{figure}

\newpage
\section*{Supplementary Note 5: ORB-v2 universal model - prediction of physical properties}

\begin{figure}[ht]
    \centering
    \includegraphics{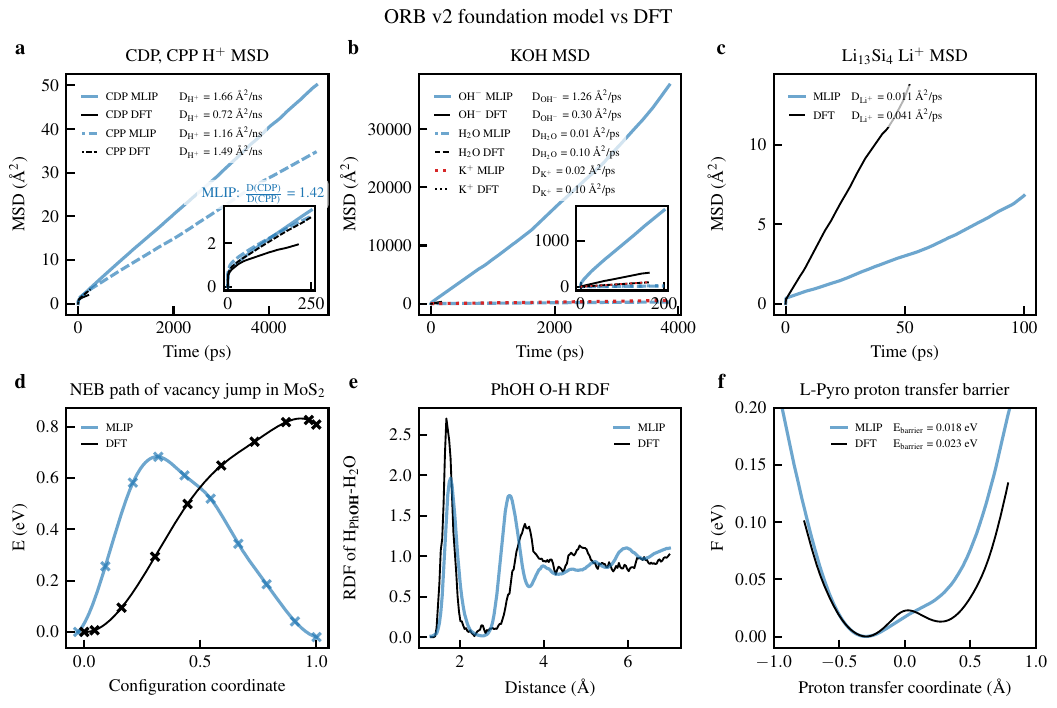}
    \caption{\textbf{ORB-V2 universal model}.
    Comparison of different physical properties obtained with first-principles methods and the universal model: \textbf{a} CsH$_2$PO$_4$ (CDP) and Cs$_7$(H$_4$PO$_4$)(H$_2$PO$_4$)$_8$ (CPP) - proton diffusion coefficients ratios of D(CDP)/D(CPP), \textbf{b} KOH - water, water molecule and hydroxide ion mean-squared displacements and diffusion coefficients, \textbf{c} Li$_{13}$Si$_4$ - lithium ion mean-squared displacements and diffusion coefficients, \textbf{d} MoS$_2$ - potential energy curves for a sulfur jump into a neighboring line of sulfur vacancies, \textbf{e} phenol in water - (H$_2$O)$\cdots$O\textsubscript{Hydroxyl-Group} radial distribution function and \textbf{f} L-pyroglutamate-ammonium (L-Pyro) - free energy profiles along the proton transfer coordinate.
    } 
\end{figure}

\newpage
\section*{Supplementary Note 6: MACE-OMAT-0 (medium) universal model - prediction of physical properties}

\begin{figure}[ht]
    \centering
    \includegraphics{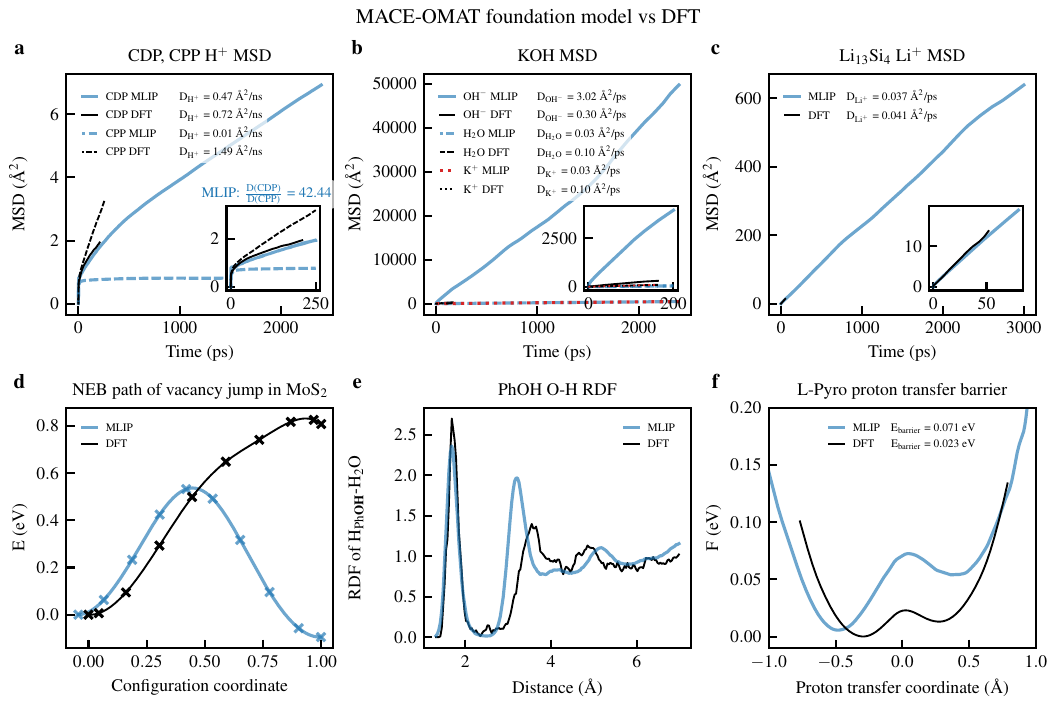}
    \caption{\textbf{MACE-OMAT-0 (medium) universal model}.
    Comparison of different physical properties obtained with first-principles methods and the universal model: \textbf{a} CsH$_2$PO$_4$ (CDP) and Cs$_7$(H$_4$PO$_4$)(H$_2$PO$_4$)$_8$ (CPP) - proton diffusion coefficients ratios of D(CDP)/D(CPP), \textbf{b} KOH - water, water molecule and hydroxide ion mean-squared displacements and diffusion coefficients, \textbf{c} Li$_{13}$Si$_4$ - lithium ion mean-squared displacements and diffusion coefficients, \textbf{d} MoS$_2$ - potential energy curves for a sulfur jump into a neighboring line of sulfur vacancies, \textbf{e} phenol in water - (H$_2$O)$\cdots$O\textsubscript{Hydroxyl-Group} radial distribution function and \textbf{f} L-pyroglutamate-ammonium (L-Pyro) - free energy profiles along the proton transfer coordinate.
    } 
\end{figure}

\newpage
\section*{Supplementary Note 7: GRACE-2L-OAM universal model - prediction of physical properties}

\begin{figure}[ht]
    \centering
    \includegraphics{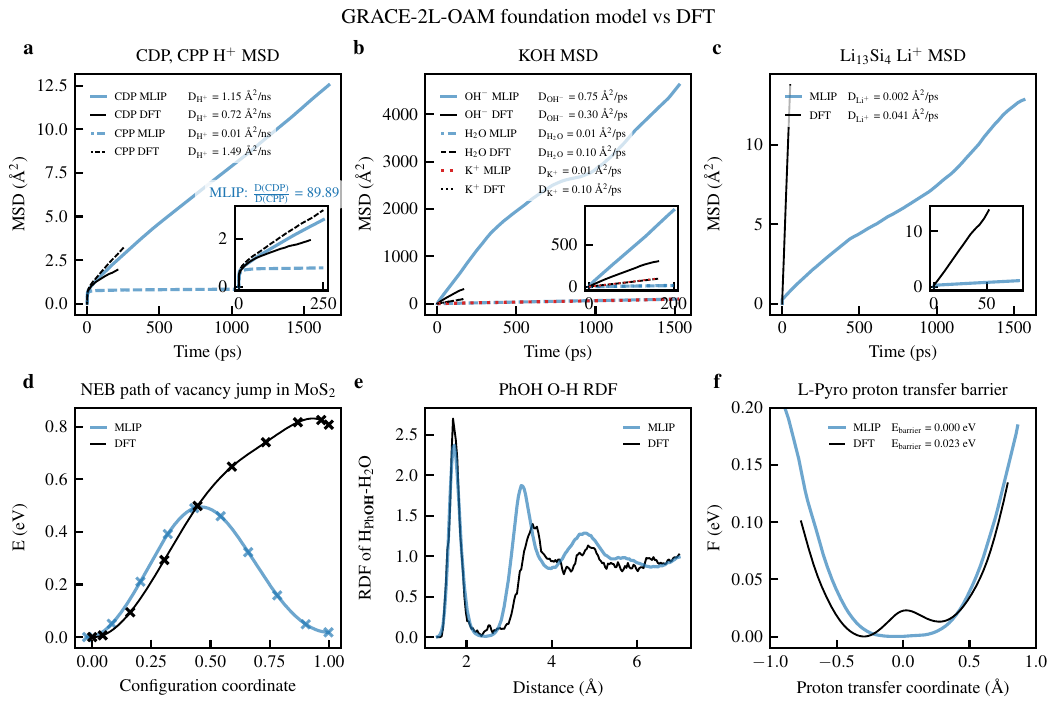}
    \caption{\textbf{GRACE-2L-OAM universal model}.
    Comparison of different physical properties obtained with first-principles methods and the universal model: \textbf{a} CsH$_2$PO$_4$ (CDP) and Cs$_7$(H$_4$PO$_4$)(H$_2$PO$_4$)$_8$ (CPP) - proton diffusion coefficients ratios of D(CDP)/D(CPP), \textbf{b} KOH - water, water molecule and hydroxide ion mean-squared displacements and diffusion coefficients, \textbf{c} Li$_{13}$Si$_4$ - lithium ion mean-squared displacements and diffusion coefficients, \textbf{d} MoS$_2$ - potential energy curves for a sulfur jump into a neighboring line of sulfur vacancies, \textbf{e} phenol in water - (H$_2$O)$\cdots$O\textsubscript{Hydroxyl-Group} radial distribution function and \textbf{f} L-pyroglutamate-ammonium (L-Pyro) - free energy profiles along the proton transfer coordinate.
    } 
\end{figure}

\newpage
\section*{Supplementary Note 8: PET-OAM-XL universal model - prediction of physical properties}

\begin{figure}[ht]
    \centering
    \includegraphics{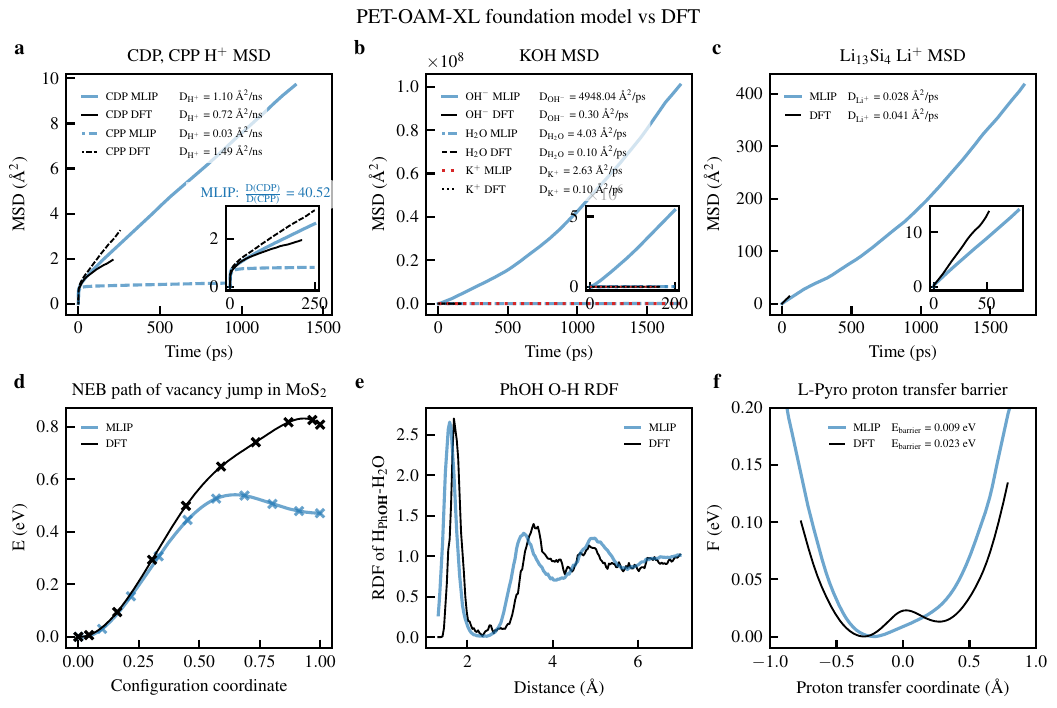}
    \caption{\textbf{PET-OAM-XL universal model}.
    Comparison of different physical properties obtained with first-principles methods and the universal model: \textbf{a} CsH$_2$PO$_4$ (CDP) and Cs$_7$(H$_4$PO$_4$)(H$_2$PO$_4$)$_8$ (CPP) - proton diffusion coefficients ratios of D(CDP)/D(CPP), \textbf{b} KOH - water, water molecule and hydroxide ion mean-squared displacements and diffusion coefficients, \textbf{c} Li$_{13}$Si$_4$ - lithium ion mean-squared displacements and diffusion coefficients, \textbf{d} MoS$_2$ - potential energy curves for a sulfur jump into a neighboring line of sulfur vacancies, \textbf{e} phenol in water - (H$_2$O)$\cdots$O\textsubscript{Hydroxyl-Group} radial distribution function and \textbf{f} L-pyroglutamate-ammonium (L-Pyro) - free energy profiles along the proton transfer coordinate.
    } 
\end{figure}

\newpage
\section*{Supplementary Note 9: PET-MAD-S-v1.5.0 universal model - prediction of physical properties}

\begin{figure}[ht]
    \centering
    \includegraphics{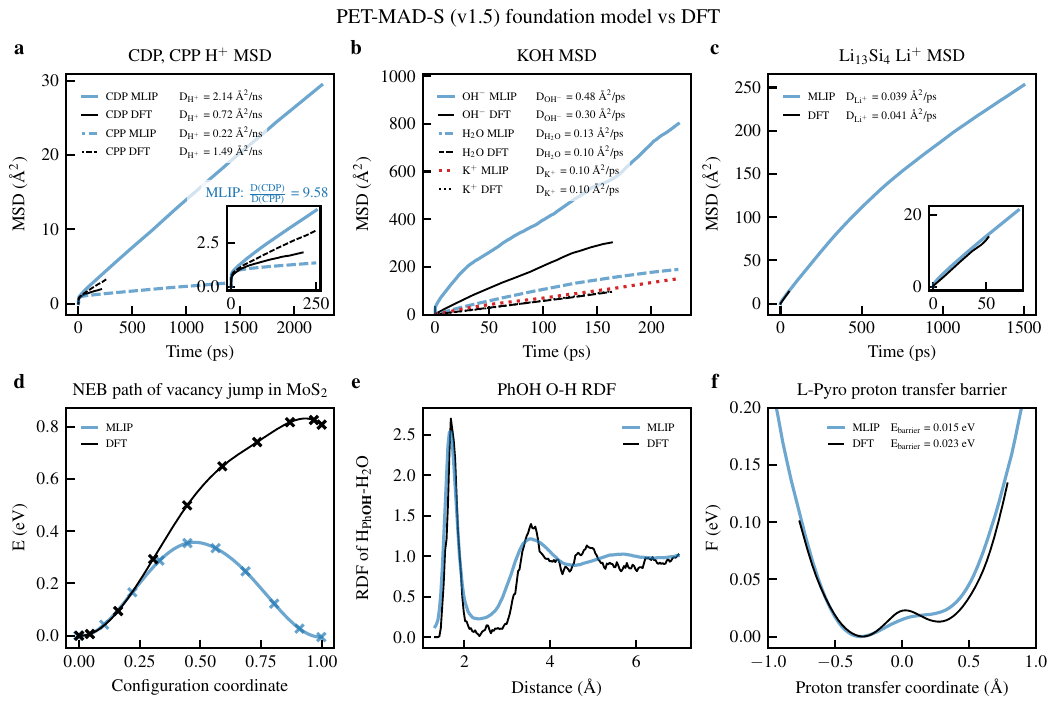}
    \caption{\textbf{PET-MAD-S-v1.5.0 universal model}.
    Comparison of different physical properties obtained with first-principles methods and the universal model: \textbf{a} CsH$_2$PO$_4$ (CDP) and Cs$_7$(H$_4$PO$_4$)(H$_2$PO$_4$)$_8$ (CPP) - proton diffusion coefficients ratios of D(CDP)/D(CPP), \textbf{b} KOH - water, water molecule and hydroxide ion mean-squared displacements and diffusion coefficients, \textbf{c} Li$_{13}$Si$_4$ - lithium ion mean-squared displacements and diffusion coefficients, \textbf{d} MoS$_2$ - potential energy curves for a sulfur jump into a neighboring line of sulfur vacancies, \textbf{e} phenol in water - (H$_2$O)$\cdots$O\textsubscript{Hydroxyl-Group} radial distribution function and \textbf{f} L-pyroglutamate-ammonium (L-Pyro) - free energy profiles along the proton transfer coordinate.
    } 
\end{figure}

\newpage
\section*{Supplementary Note 10: Universal model performance - Force errors}

\begin{figure}[ht]
    \centering
    \includegraphics{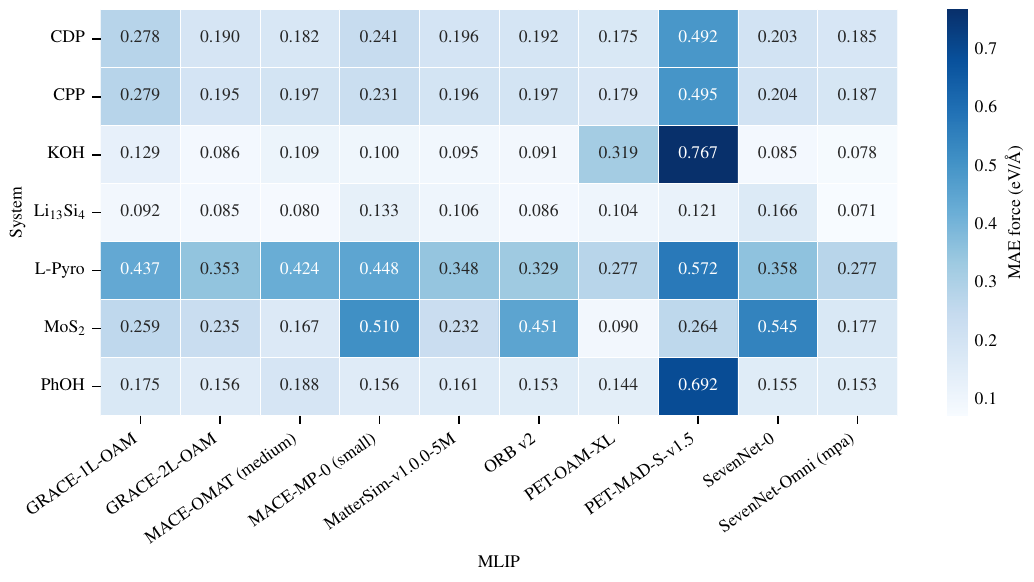}
    \caption{\textbf{Universal model force errors}.
    Mean-absolute-error on the predicted forces (in eV/\AA) of the different universal MLIPs on the different systems: CsH$_2$PO$_4$ (CDP) and Cs$_7$(H$_4$PO$_4$)(H$_2$PO$_4$)$_8$ (CPP), aqueous KOH solution, Li$_{13}$Si$_4$, MoS$_2$, phenol in water and L-pyroglutamate-ammonium (L-Pyro).
    } 
\end{figure}

\newpage
\section*{Supplementary Note 11: Universal model performance - Energy errors}

\begin{figure}[ht]
    \centering
    \includegraphics{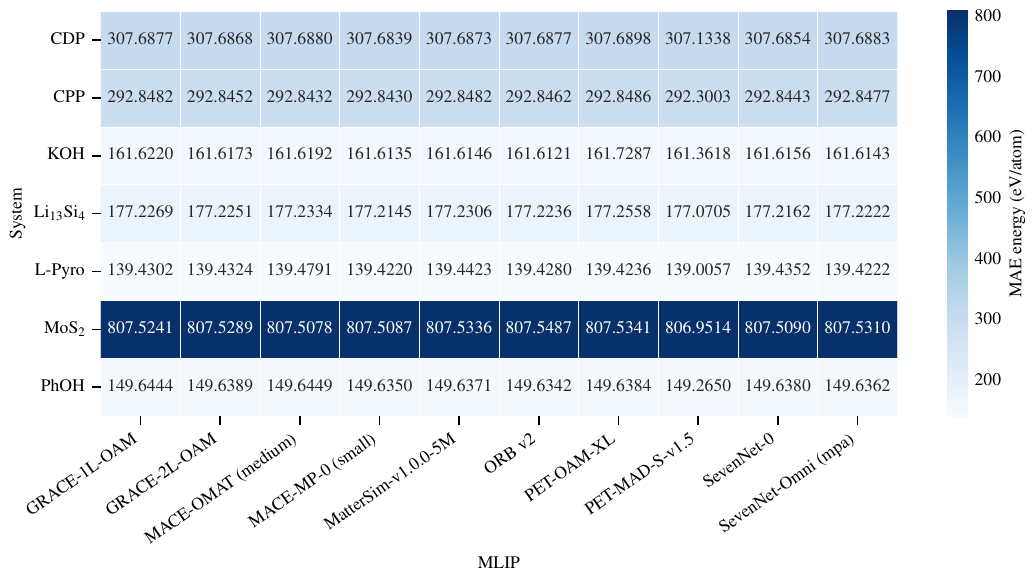}
    \caption{\textbf{Universal model energy errors}.
    Mean-absolute-error on the predicted energies of the different universal MLIPs on the different systems: CsH$_2$PO$_4$ (CDP) and Cs$_7$(H$_4$PO$_4$)(H$_2$PO$_4$)$_8$ (CPP), aqueous KOH solution, Li$_{13}$Si$_4$, MoS$_2$, phenol in water and L-pyroglutamate-ammonium (L-Pyro).
    } 
\end{figure}

\newpage
\section*{Supplementary Note 12: Discussion of universal model performances}

A detailed discussion of the performance of the earlier universal models, namely \textsc{MACE-MP-0} (small) \cite{mace_mp}, \textsc{MatterSim-v1.0.0-5M} \cite{mattersim}, \textsc{SevenNet-0} \cite{sevennet_1}, \textsc{GRACE-1L-OAM} \cite{grace_2}, and \textsc{ORB-v2} \cite{orbv2}, in predicting the physical properties of the seven investigated systems is provided in our previous work \cite{haenseroth2026atk}. 
These results are summarized in Supplementary Notes~1-5 and 11-12.

The performance of four newer universal models, namely \textsc{MACE-OMAT-0} (medium) \cite{batatia2025cross}, \textsc{GRACE-2L-OAM} \cite{grace_2}, \textsc{PET-MAD-S-v1.5.0} \cite{pet_mad, pet_mad_15}, and \textsc{PET-OAM-XL} \cite{pet_oam}, in predicting the material-specific properties across the seven systems is shown in Supplementary Notes~6-12. 

The force errors reported in Supplementary Note~11 show that these newer universal models perform substantially better than the previously investigated models. 
For several materials, some of the models achieve force errors below $0.1$~eV/{\AA}. 
Among all investigated models, \textsc{PET-OAM-XL} yields the lowest overall errors, achieving the best performance for five of the seven systems.

\textsc{PET-OAM-XL} also shows the best performance in predicting the energy profile of the sulfur-vacancy jump in MoS$_2$, although it still does not reach DFT-level accuracy. 
Furthermore, it is able to reproduce the solvation structure of phenol in water and the lithium-ion diffusion in Li$_{13}$Si$_4$. 
These properties are also well described by the \textsc{MACE-OMAT-0} (medium) universal model.

The \textsc{PET-MAD-S-v1.5.0} model is able to predict the solvation of the OH group of phenol in water, as well as Li$^+$ diffusion in Li$_{13}$Si$_4$ and the mobility of ions and molecules in aqueous potassium hydroxide solution.
The \textsc{GRACE-2L-OAM} model reproduces the local solvation environment of the phenol OH group in water and the mobility of K$^+$ and H$_2$O in aqueous potassium hydroxide solution, but overestimates the hydroxide-ion diffusion.

Overall, these newer models show improved zero-shot accuracy compared with the previously investigated universal models. This improvement is likely related to their larger and more diverse training datasets, which include data from \textsc{Alexandria} \cite{alexandria,omat24}, the \textsc{Materials Project} \cite{mp_1,mp_2,mptrj}, \textsc{Open Materials 2024} \cite{omat24}, and \textsc{Open Molecules 2025} \cite{omol25}, as well as to the increased number of parameters in these models.

\newpage
\section*{Supplementary Note 13: Fine-tuned MACE model performance - Energy errors - Dataset generating universal model vs. systems vs. dataset size (universal model dataset)}

\begin{figure}[ht]
    \centering
    \includegraphics{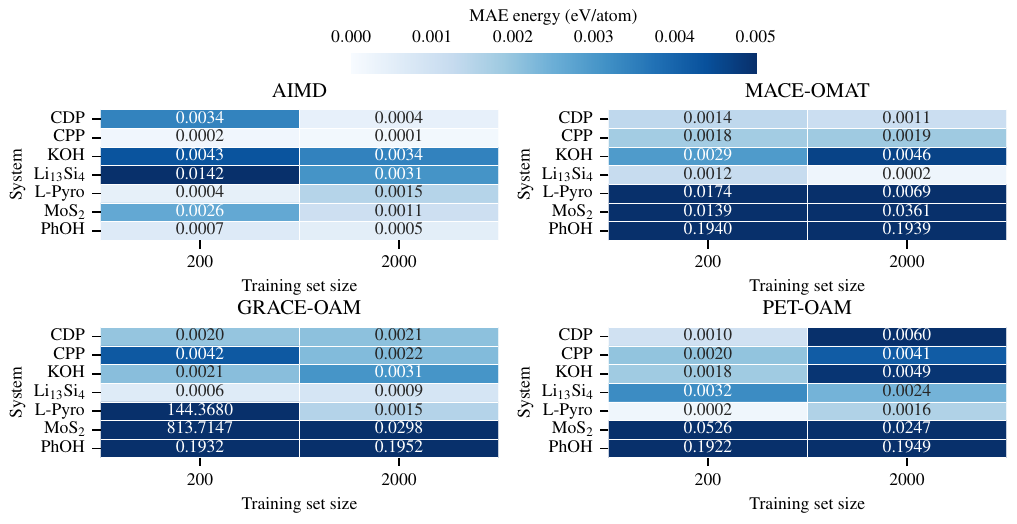}
    \caption{\textbf{Fine-tuned MACE model energy errors}.
    Mean-absolute-error on the predicted energies of the fine-tuned MACE-MP-0 universal MLIPs (different universal dataset, 2000 frames) on the different systems: CsH$_2$PO$_4$ (CDP) and Cs$_7$(H$_4$PO$_4$)(H$_2$PO$_4$)$_8$ (CPP), aqueous KOH solution, Li$_{13}$Si$_4$, MoS$_2$, phenol in water and L-pyroglutamate-ammonium (L-Pyro).
    } 
\end{figure}

\newpage
\section*{Supplementary Note 14: Fine-tuned model (universal model dataset) - prediction of potential energy profile for sulfur-vacancy jump in MoS$_2$}

\begin{figure}[ht]
    \centering
    \includegraphics{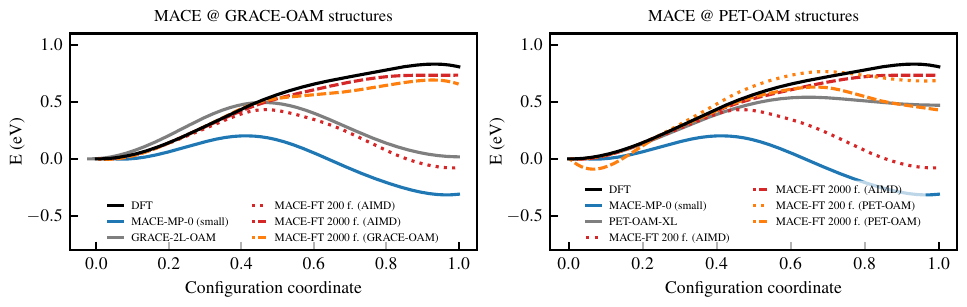}
    \caption{\textbf{Energy profiles computed with fine-tuned models with variable universal model dataset}.
    Potential energy curves for a sulfur jump into a neighboring line of sulfur vacancies in MoS$_2$ via NEB calculations using fine-tuned \textsc{MACE-MP-0} (small) universal models on DFT-recalculated universal model structures (orange lines) obtained with \textsc{GRACE-2L-OAM} (left) and \textsc{PET-OAM-XL} (right); and fine-tuned \textsc{MACE-MP-0} (small) universal models on sub-sampled AIMD trajectories (red lines); with dataset sizes of $200$~(dotted lines) and $2000$~(dashed lines), \textsc{MACE-MP-0} (small) (solid blue line) and the respective universal model (solid gray line) as well as the DFT reference profile (solid black line).
    The model fine-tuned on the dataset consisting out of $200$~recalculated \textsc{GRACE-2L-OAM} data points is not able to compute converged structures for the NEB replicas.
    } 
\end{figure}

\newpage
\section*{Supplementary Note 15: MACE-MP-0 fine-tuned universal model (MACE-OMAT dataset) - prediction of physical properties}

\begin{figure}[ht]
    \centering
    \includegraphics{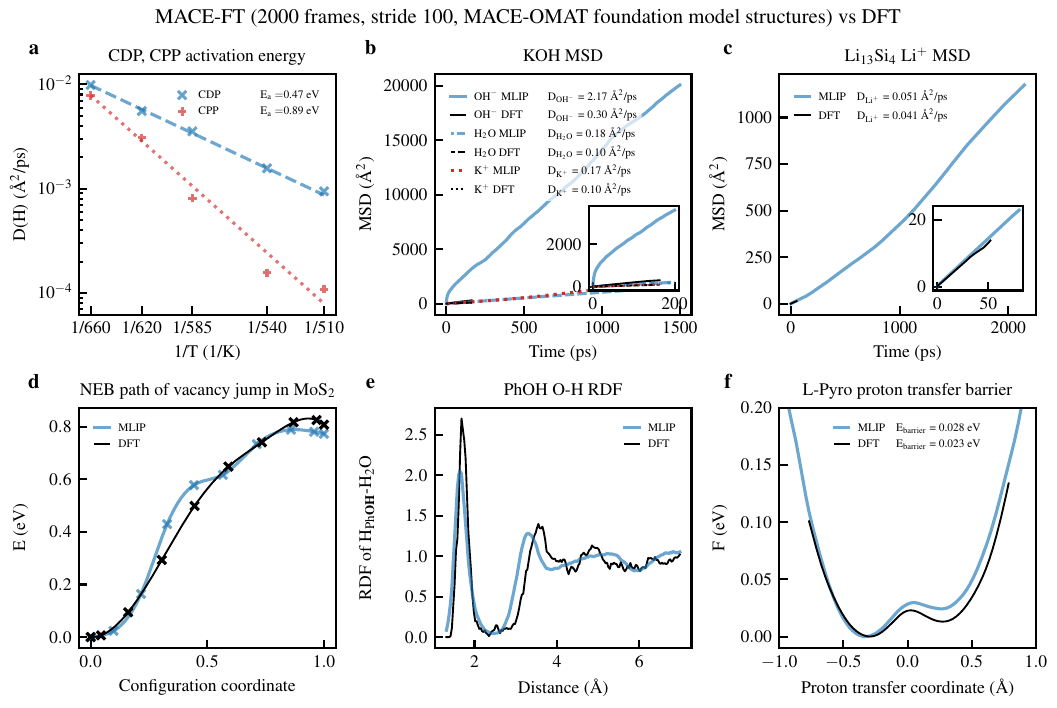}
    \caption{\textbf{MACE-FT finetuned universal model on MACE-OMAT dataset}.
    Comparison of different physical properties obtained with first-principles methods and the fine-tuned universal models (MACE-OMAT data, 2000 frames, stride 3000): \textbf{a} CsH$_2$PO$_4$ (CDP) and Cs$_7$(H$_4$PO$_4$)(H$_2$PO$_4$)$_8$ (CPP) - activation energy of proton diffusion, \textbf{b} KOH - water, water molecule and hydroxide ion mean-squared displacements and diffusion coefficients, \textbf{c} Li$_{13}$Si$_4$ - lithium ion mean-squared displacements and diffusion coefficients, \textbf{d} MoS$_2$ - potential energy curves for a sulfur jump into a neighboring line of sulfur vacancies, \textbf{e} phenol in water - (H$_2$O)$\cdots$O\textsubscript{Hydroxyl-Group} radial distribution function and \textbf{f} L-pyroglutamate-ammonium (L-Pyro) - free energy profiles along the proton transfer coordinate.
    } 
\end{figure}

\newpage
\section*{Supplementary Note 16: MACE-MP-0 fine-tuned universal model (PET-OAM dataset) - prediction of physical properties}

\begin{figure}[ht]
    \centering
    \includegraphics{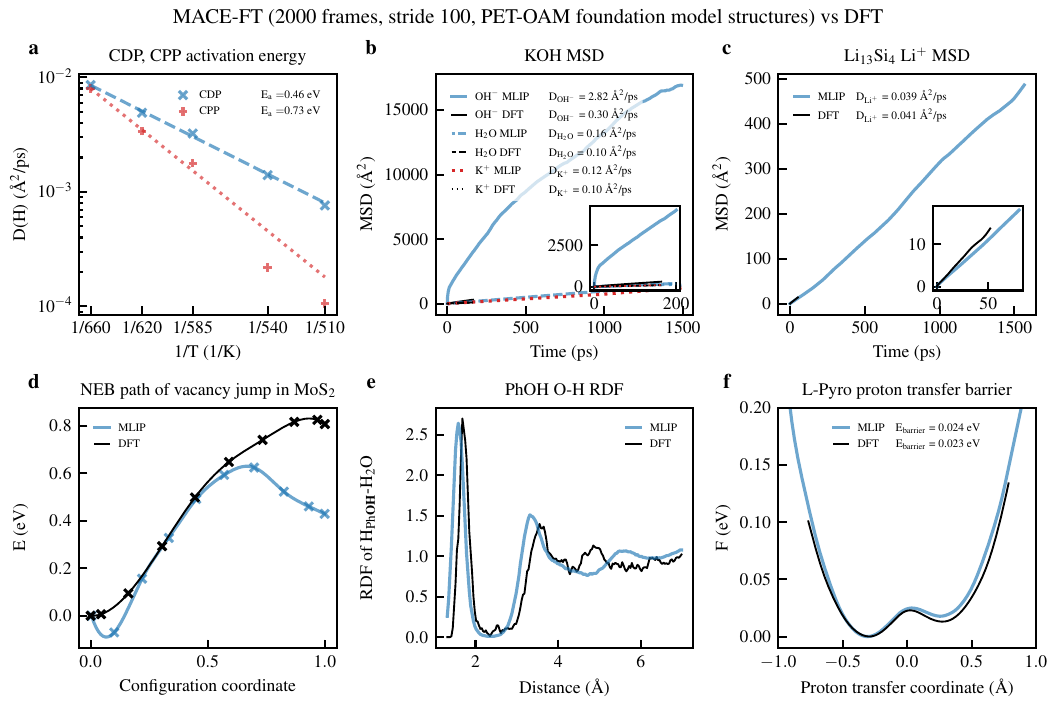}
    \caption{\textbf{MACE finetuned universal model on PET-OAM dataset}.
    Comparison of different physical properties obtained with first-principles methods and the fine-tuned universal models (PET-OAM data, 2000 frames, stride 3000): \textbf{a} CsH$_2$PO$_4$ (CDP) and Cs$_7$(H$_4$PO$_4$)(H$_2$PO$_4$)$_8$ (CPP) - activation energy of proton diffusion, \textbf{b} KOH - water, water molecule and hydroxide ion mean-squared displacements and diffusion coefficients, \textbf{c} Li$_{13}$Si$_4$ - lithium ion mean-squared displacements and diffusion coefficients, \textbf{d} MoS$_2$ - potential energy curves for a sulfur jump into a neighboring line of sulfur vacancies, \textbf{e} phenol in water - (H$_2$O)$\cdots$O\textsubscript{Hydroxyl-Group} radial distribution function and \textbf{f} L-pyroglutamate-ammonium (L-Pyro) - free energy profiles along the proton transfer coordinate.
    } 
\end{figure}

\newpage
\section*{Supplementary Note 17: MACE-MP-0 fine-tuned universal model (AIMD dataset) - prediction of physical properties}

\begin{figure}[ht]
    \centering
    \includegraphics{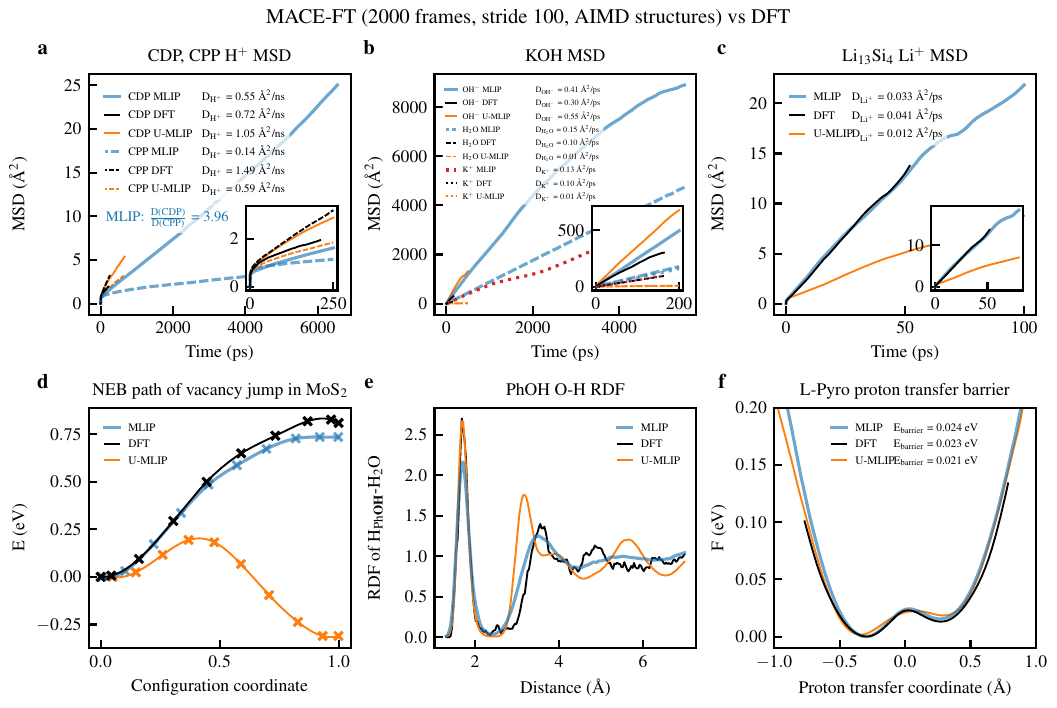} 
    \caption{\textbf{MACE-MP-0 finetuned universal model on AIMD dataset}.
    Comparison of different physical properties obtained with first-principles methods (black lines) and universal model MACE-MP-0 (orange lines) and the fine-tuned universal models (blue line, AIMD dataset, 2000 frames, stride 100): \textbf{a} CsH$_2$PO$_4$ (CDP) and Cs$_7$(H$_4$PO$_4$)(H$_2$PO$_4$)$_8$ (CPP) - activation energy of proton diffusion, \textbf{b} KOH - water, water molecule and hydroxide ion mean-squared displacements and diffusion coefficients, \textbf{c} Li$_{13}$Si$_4$ - lithium ion mean-squared displacements and diffusion coefficients, \textbf{d} MoS$_2$ - potential energy curves for a sulfur jump into a neighboring line of sulfur vacancies, \textbf{e} phenol in water - (H$_2$O)$\cdots$O\textsubscript{Hydroxyl-Group} radial distribution function and \textbf{f} L-pyroglutamate-ammonium (L-Pyro) - free energy profiles along the proton transfer coordinate.
    } 
\end{figure}

\newpage
\section*{Supplementary Note 18: Fine-tuned model performance - Force errors - MLIPs vs. systems vs. dataset size (MACE-OMAT universal model dataset)}

\begin{figure}[ht]
    \centering
    \includegraphics{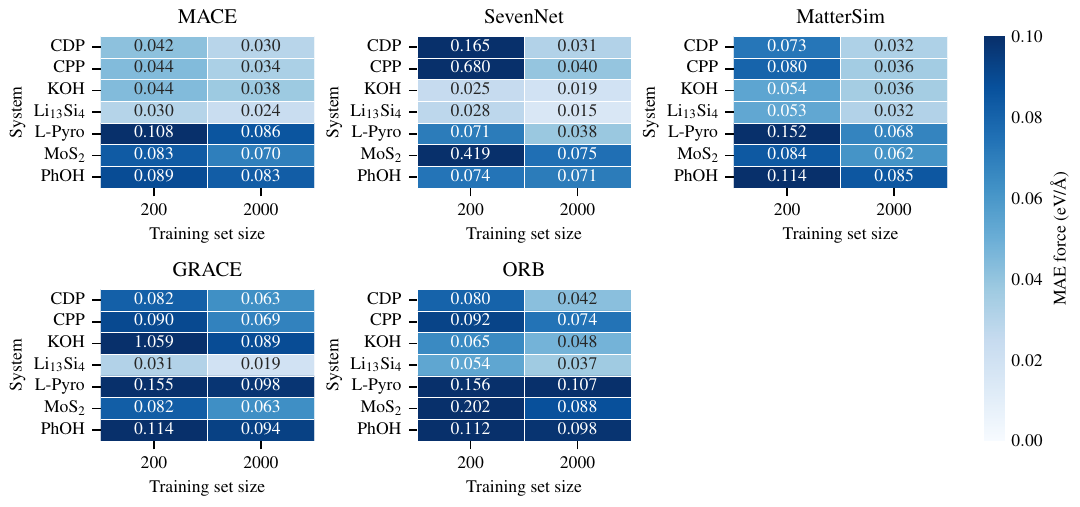}
    \caption{\textbf{Fine-tuned model force errors}.
    Mean-absolute-error on the predicted forces of the different fine-tuned MLIPs (same MACE-OMAT-0 (medium) universal dataset, 2000 frames) on the different systems: CsH$_2$PO$_4$ (CDP) and Cs$_7$(H$_4$PO$_4$)(H$_2$PO$_4$)$_8$ (CPP), aqueous KOH solution, Li$_{13}$Si$_4$, MoS$_2$, phenol in water and L-pyroglutamate-ammonium (L-Pyro).
    } 
\end{figure}

\newpage
\section*{Supplementary Note 19: Fine-tuned model performance - Energy errors - MLIPs vs. systems vs. dataset size (MACE-OMAT universal model dataset)}

\begin{figure}[ht]
    \centering
    \includegraphics{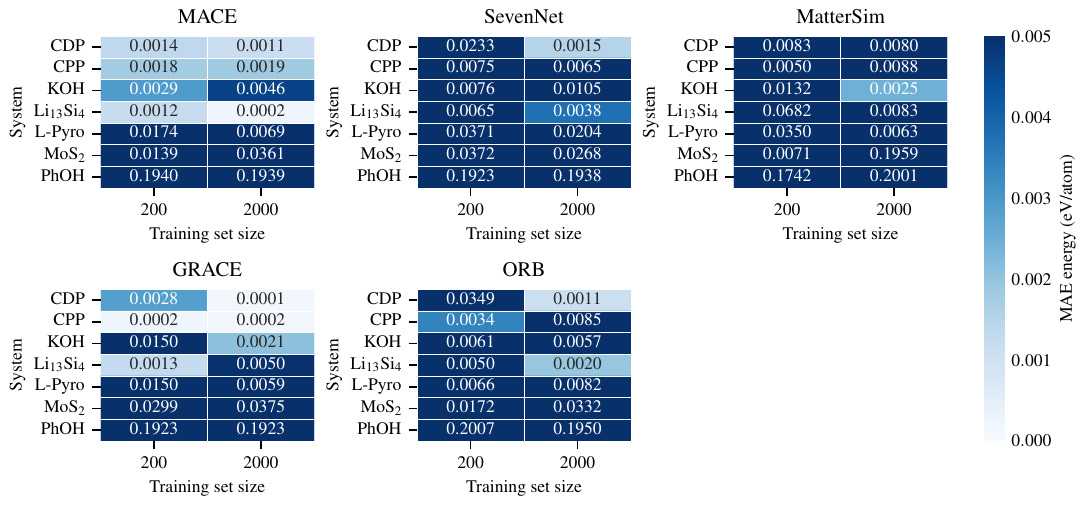}
    \caption{\textbf{Fine-tuned model energy errors}.
    Mean-absolute-error on the predicted energies of the different fine-tuned MLIPs (same \textsc{MACE-OMAT-0} (medium) universal dataset, 2000 frames) on the different systems: CsH$_2$PO$_4$ (CDP) and Cs$_7$(H$_4$PO$_4$)(H$_2$PO$_4$)$_8$ (CPP), aqueous KOH solution, Li$_{13}$Si$_4$, MoS$_2$, phenol in water and L-pyroglutamate-ammonium (L-Pyro).
    } 
\end{figure}

\newpage
\section*{Supplementary Note 20: Fine-tuned model (\textsc{MACE-OMAT-0} model dataset) - prediction of potential energy profile for sulfur-vacancy jump in MoS$_2$}

\begin{figure}[ht]
    \centering
    \includegraphics{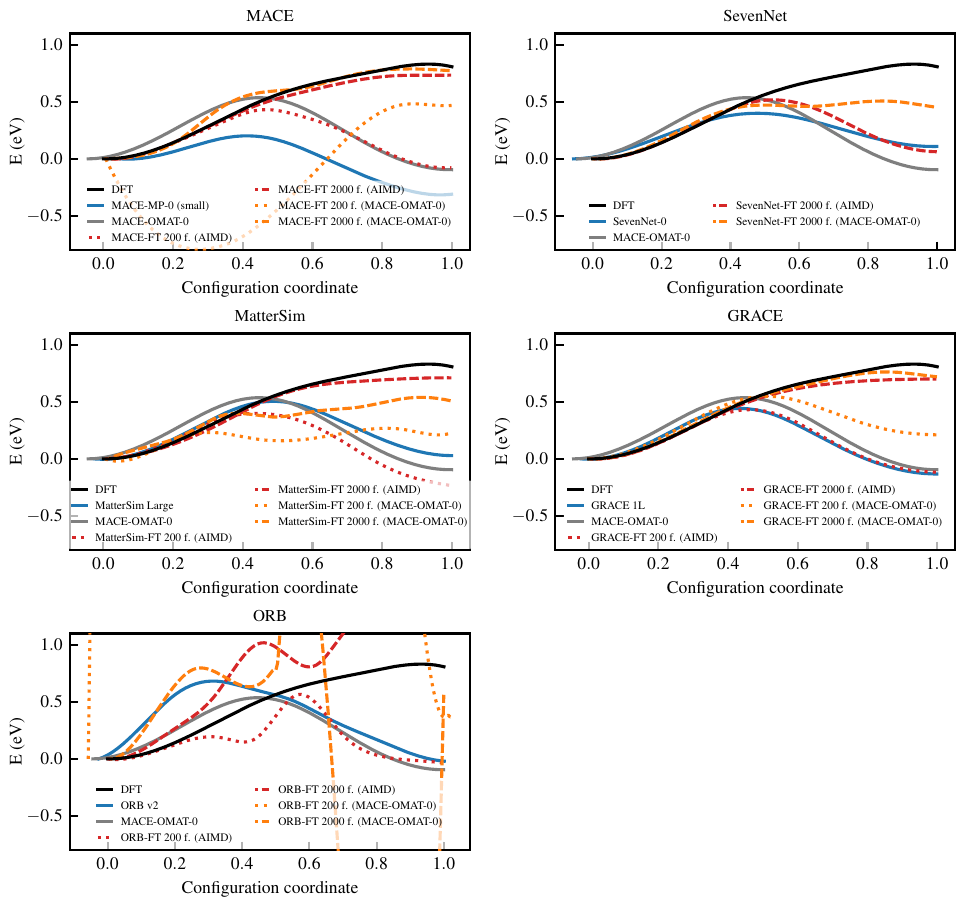}
    \caption{\textbf{Energy profiles computed with fine-tuned models with variable universal model dataset}.
    Potential energy curves for a sulfur jump into a neighboring line of sulfur vacancies in MoS$_2$ via NEB calculations using fine-tuned \textsc{MACE-MP-0} (small), \textsc{SevenNet-0}, \textsc{MatterSim-v1.0.0-5M}, \textsc{GRACE-1L-OAM} and \textsc{ORB-v2} universal models on DFT-recalculated universal model structures (orange lines) obtained with \textsc{MACE-OMAT-0} (medium); and fine-tuned universal models on sub-sampled AIMD trajectories (red lines); with dataset sizes of $200$~(dotted lines) and $2000$~(dashed lines), the respective universal model (solid blue line) and \textsc{MACE-OMAT-0} (medium)(solid gray line) as well as the DFT reference profile (solid black line).
    The universal \textsc{SevenNet-0} models fine-tuned on the datasets consisting out of $200$~data points sub-sampled from an AIMD trajectories and $200$~recalculated \textsc{MACE-OMAT-0} (medium) data points are not able to compute converged structures for the NEB replicas.
    } 
\end{figure}

\newpage
\section*{Supplementary Note 21: SevenNet-0 fine-tuned universal model (MACE-OMAT dataset) - prediction of physical properties}

\begin{figure}[ht]
    \centering
    \includegraphics{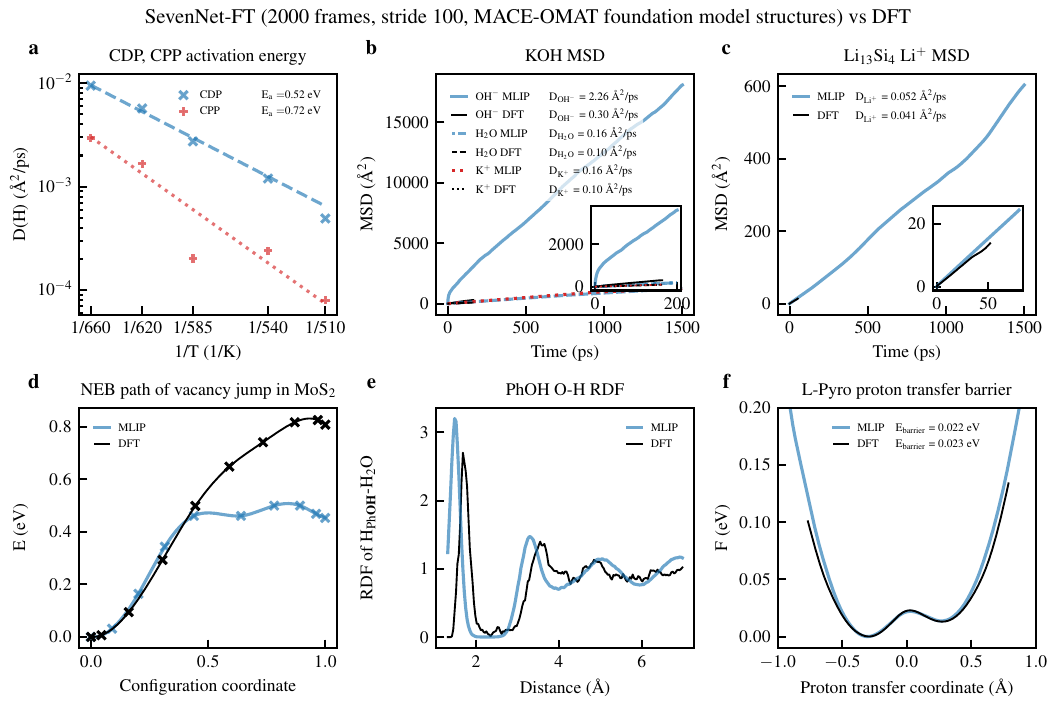}
    \caption{\textbf{SevenNet-FT finetuned universal model on MACE-OMAT dataset}.
    Comparison of different physical properties obtained with first-principles methods and the fine-tuned universal models (MACE-OMAT data, 2000 frames, stride 3000): \textbf{a} CsH$_2$PO$_4$ (CDP) and Cs$_7$(H$_4$PO$_4$)(H$_2$PO$_4$)$_8$ (CPP) - activation energy of proton diffusion, \textbf{b} KOH - water, water molecule and hydroxide ion mean-squared displacements and diffusion coefficients, \textbf{c} Li$_{13}$Si$_4$ - lithium ion mean-squared displacements and diffusion coefficients, \textbf{d} MoS$_2$ - potential energy curves for a sulfur jump into a neighboring line of sulfur vacancies, \textbf{e} phenol in water - (H$_2$O)$\cdots$O\textsubscript{Hydroxyl-Group} radial distribution function and \textbf{f} L-pyroglutamate-ammonium (L-Pyro) - free energy profiles along the proton transfer coordinate.
    } 
\end{figure}

\newpage
\section*{Supplementary Note 22: MatterSim-Large fine-tuned universal model (MACE-OMAT dataset) - prediction of physical properties}

\begin{figure}[ht]
    \centering
    \includegraphics{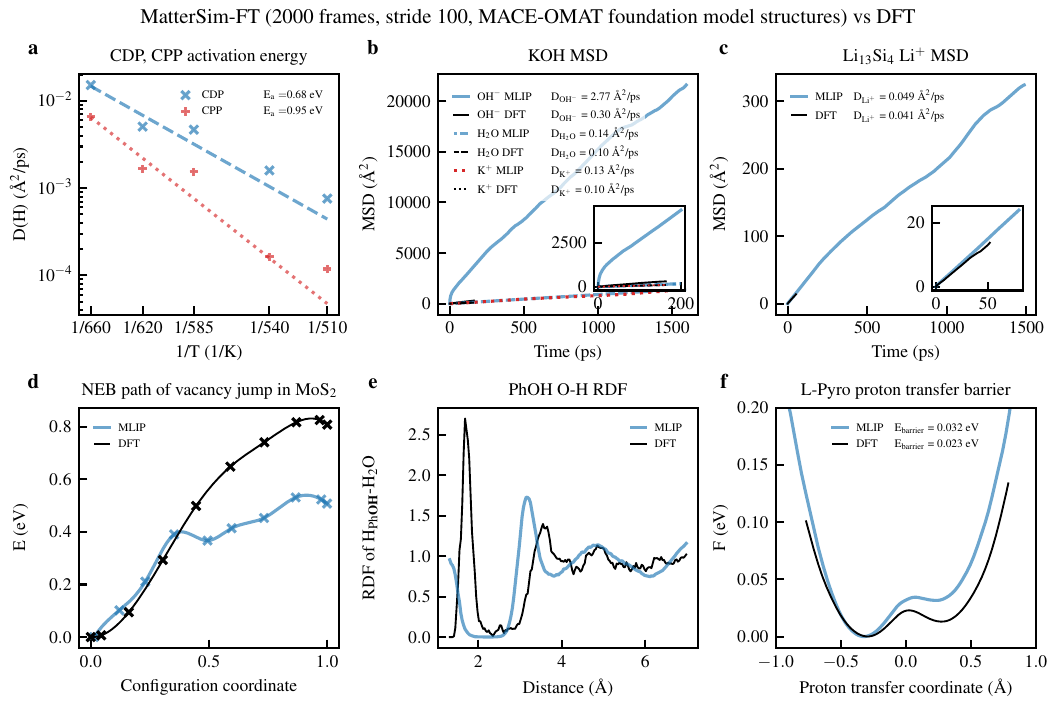}
    \caption{\textbf{MatterSim-FT finetuned universal model on MACE-OMAT dataset}.
    Comparison of different physical properties obtained with first-principles methods and the fine-tuned universal models (MACE-OMAT data, 2000 frames, stride 3000): \textbf{a} CsH$_2$PO$_4$ (CDP) and Cs$_7$(H$_4$PO$_4$)(H$_2$PO$_4$)$_8$ (CPP) - activation energy of proton diffusion, \textbf{b} KOH - water, water molecule and hydroxide ion mean-squared displacements and diffusion coefficients, \textbf{c} Li$_{13}$Si$_4$ - lithium ion mean-squared displacements and diffusion coefficients, \textbf{d} MoS$_2$ - potential energy curves for a sulfur jump into a neighboring line of sulfur vacancies, \textbf{e} phenol in water - (H$_2$O)$\cdots$O\textsubscript{Hydroxyl-Group} radial distribution function and \textbf{f} L-pyroglutamate-ammonium (L-Pyro) - free energy profiles along the proton transfer coordinate.
    } 
\end{figure}

\newpage
\section*{Supplementary Note 23: Trained MACE model performance - Force errors - Dataset generating universal model vs. systems vs. dataset size (universal model dataset)}

\begin{figure}[ht]
    \centering
    \includegraphics{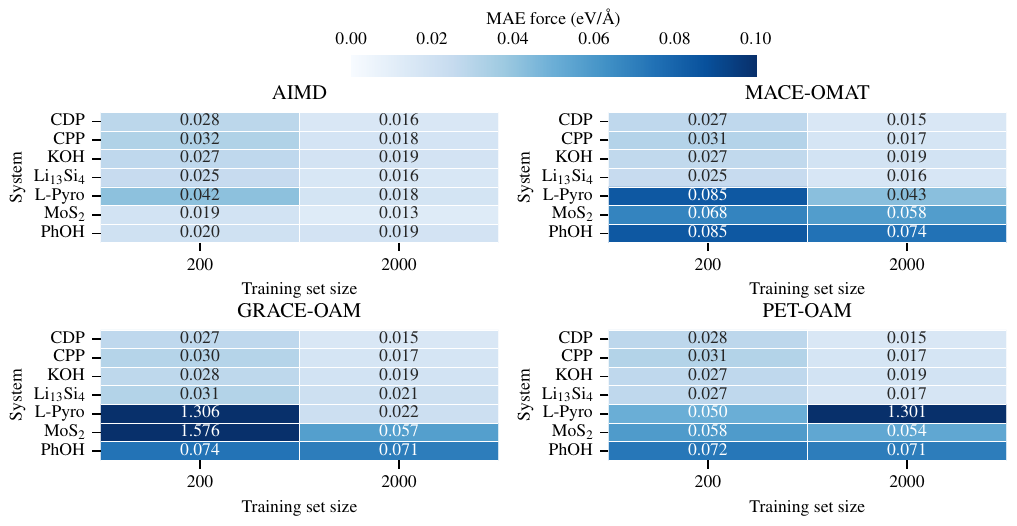}
    \caption{\textbf{Trained-from-scratch MACE model force errors}.
    Mean-absolute-error on the predicted forces of the trained-from-scratch MACE MLIPs (different universal dataset, 2000 frames) on the different systems: CsH$_2$PO$_4$ (CDP) and Cs$_7$(H$_4$PO$_4$)(H$_2$PO$_4$)$_8$ (CPP), aqueous KOH solution, Li$_{13}$Si$_4$, MoS$_2$, phenol in water and L-pyroglutamate-ammonium (L-Pyro).
    } 
\end{figure}

\newpage
\section*{Supplementary Note 24: Trained MACE model performance - Energy errors - Dataset generating universal model vs. systems vs. dataset size (universal model dataset)}

\begin{figure}[ht]
    \centering
    \includegraphics{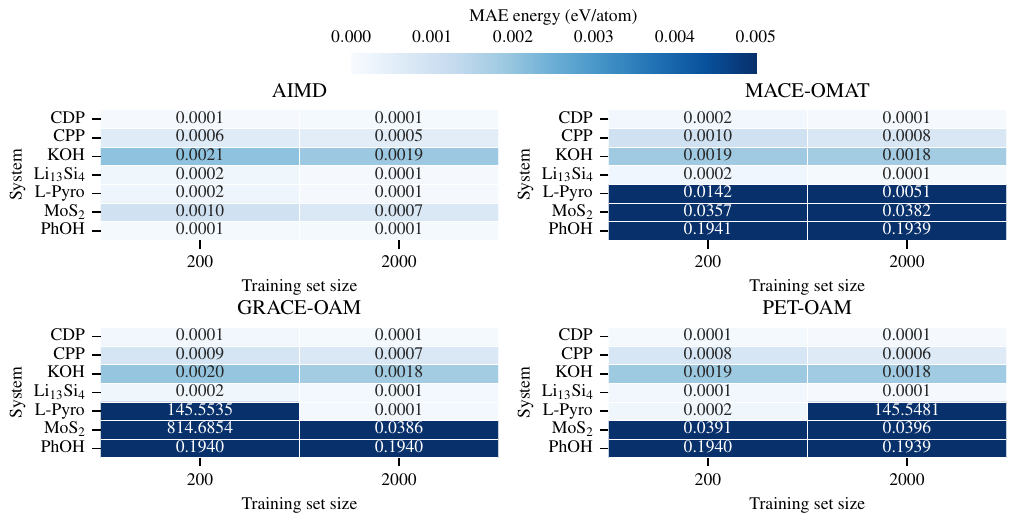}
    \caption{\textbf{Trained-from-scratch MACE model energy errors}.
    Mean-absolute-error on the predicted energies of the trained-from-scratch MACE MLIPs (different universal dataset, 2000 frames) on the different systems: CsH$_2$PO$_4$ (CDP) and Cs$_7$(H$_4$PO$_4$)(H$_2$PO$_4$)$_8$ (CPP), aqueous KOH solution, Li$_{13}$Si$_4$, MoS$_2$, phenol in water and L-pyroglutamate-ammonium (L-Pyro).
    } 
\end{figure}

\newpage
\section*{Supplementary Note 25: Trained-from-scratch model (universal model dataset) - prediction of potential energy profile for sulfur-vacancy jump in MoS$_2$}

\begin{figure}[ht]
    \centering
    \includegraphics{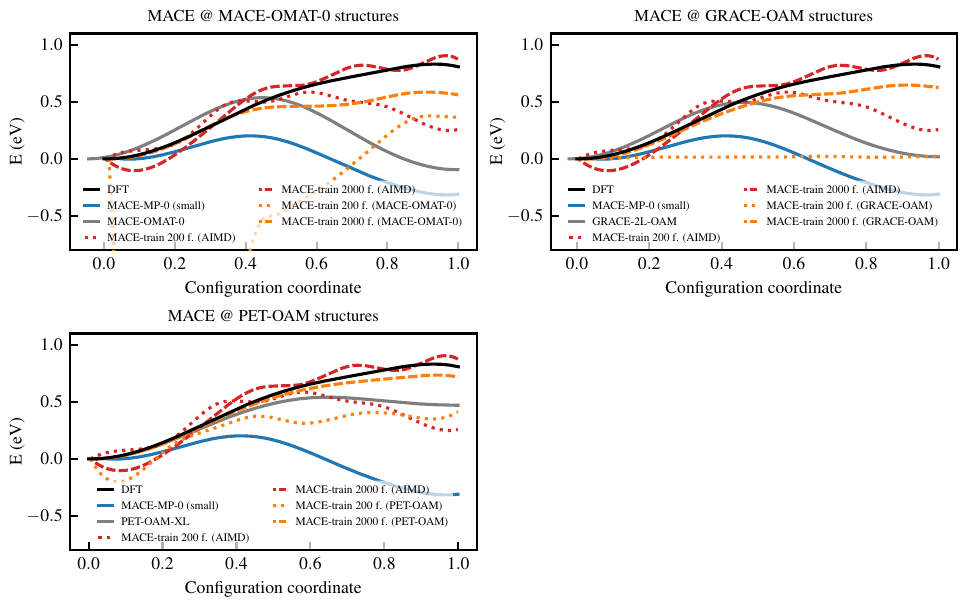}
    \caption{\textbf{Energy profiles computed with trained-from-scratch models with variable universal model dataset}.
    Potential energy curves for a sulfur jump into a neighboring line of sulfur vacancies in MoS$_2$ via NEB calculations using trained-from-scratch \textsc{MACE} models on DFT-recalculated universal model structures (orange lines) obtained with \textsc{GRACE-2L-OAM} (left) and \textsc{PET-OAM-XL} (right) and trained-from-scratch \textsc{MACE} models on sub-sampled AIMD trajectories (red lines); with dataset sizes of $200$~(dotted lines) and $2000$~(dashed lines), \textsc{MACE-MP-0} (small) (solid blue line) and the respective universal model (solid gray line) as well as the DFT reference profile (solid black line).
    } 
\end{figure}

\newpage
\section*{Supplementary Note 26: Comparing iterative training against single time fine-tuning - prediction of potential energy profile for sulfur-vacancy jump in MoS$_2$}

\begin{figure}[ht]
    \centering
    \includegraphics{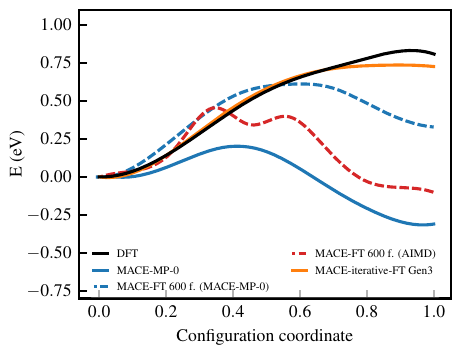}
    \caption{\textbf{Energy profiles computed with iterative-fine-tuned and singular fine-tuned models}.
    Potential energy curves for a sulfur jump into a neighboring line of sulfur vacancies in MoS$_2$ via NEB calculations using datasets build up from: $600$~data points from a sub-sampled \textsc{MACE-MP-0} trajectory (dashed blue line) and $600$~data points from a sub-sampled AIMD trajectory (dashed red line),  the third generation of the iterative training approach shown in main text in Figure~5e (solid orange line), the respective universal model (solid blue line) as well as the DFT reference profile (solid black line).
    } 
\end{figure}

\newpage
\section*{Supplementary Note 27: Iterative training additional model generations}

\begin{figure}[ht]
    \centering
    \includegraphics{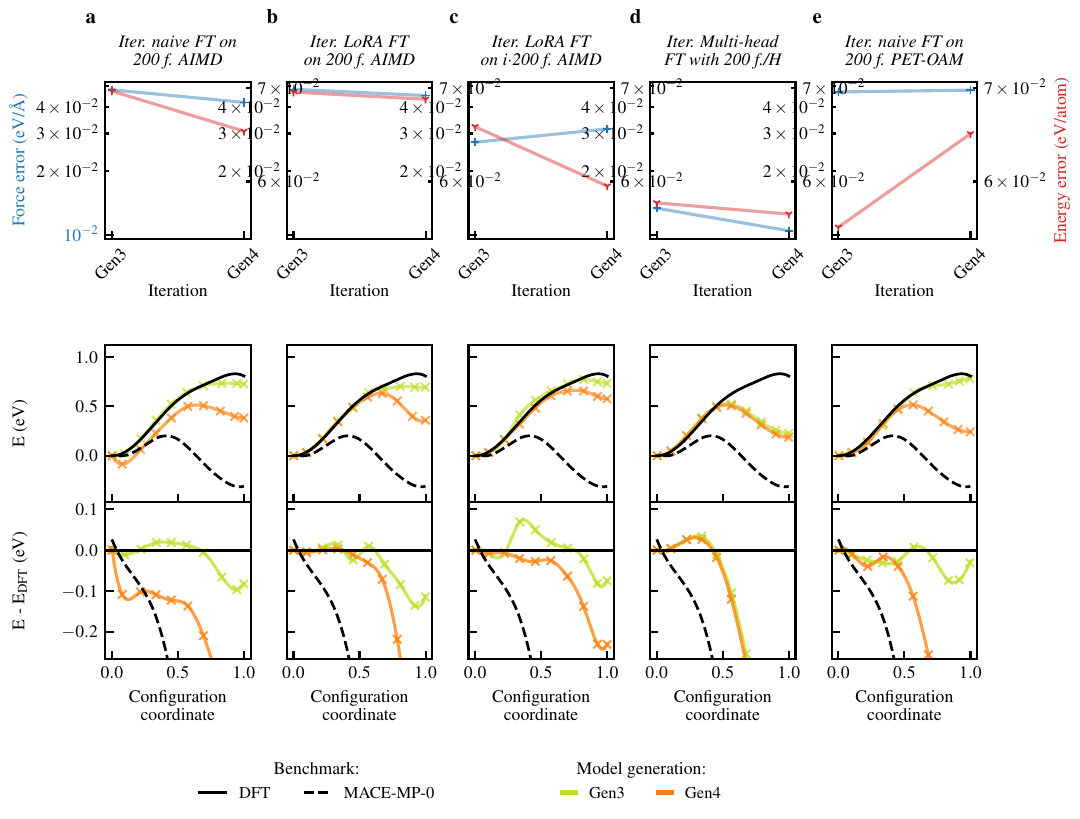}
    \caption{\textbf{Iterative model training generations}.
    Iterative training models of generation $3$ and $4$.
    The models of generation $4$ tend to perform worse then the previous generation because of the bad training data composition.
    } 
\end{figure}

\newpage
\section*{Supplementary Note 28: Comparing iterative training dataset composition}

\begin{figure}[ht]
    \centering
    \includegraphics{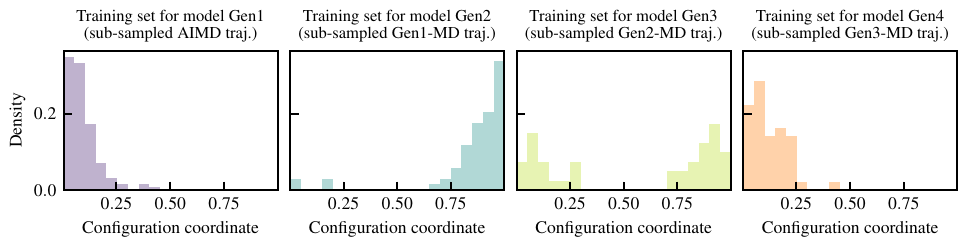}
    \caption{\textbf{Dataset composition for different model generation}.
    Training set composition with respect to the configuration coordinate of the sulfur-vacancy jump of the different model generations.
    } 
\end{figure}

\newpage
\bibliography{bibliography_si}